\documentclass[a4paper,fleqn,preprint]{cas-dc}
\usepackage{draftwatermark}  
\SetWatermarkText{PREPRINT}  
\SetWatermarkScale{0.5}      
\SetWatermarkColor[gray]{0.8} 
\usepackage[numbers]{natbib}
\usepackage{csquotes}
\usepackage{cleveref}
\usepackage{glossaries}
\usepackage{pifont}

\newglossaryentry{qos}
{
    name=todo,
    description={}
}

\usepackage{enumitem}
\newcommand{\subscript}[2]{$#1 _ #2$}

\def\tsc#1{\csdef{#1}{\textsc{\lowercase{#1}}\xspace}}
\tsc{WGM}
\tsc{QE}
\tsc{EP}
\tsc{PMS}
\tsc{BEC}
\tsc{DE}

\newtheorem{definition}{Definition}

\begin{document}
\let\WriteBookmarks\relax
\def\floatpagepagefraction{1}
\def\textpagefraction{.001}
\shorttitle{An Architectural View Type for Elasticity Modeling and Simulation}
\shortauthors{Klinaku et~al.}

\title [mode = title]{An Architectural View Type for Elasticity Modeling and Simulation\\---The Slingshot Approach}

\author[1]{Floriment Klinaku}[orcid=0000-0001-0000-0000]

\author[1]{Sarah Sophie Stieß}

\author[2]{Alireza Hakamian}

\author[1]{Steffen Becker}

\affiliation[1]{organization={ Institute of Software Engineering (ISTE), University of Stuttgart},
                city={Stuttgart},
                country={Germany}}

                \affiliation[2]{organization={
                Software Engineering and Construction Method (SWK), University of Hamburg},
                city={Hamburg},
                country={Germany}}

\cortext[cor1]{Corresponding author}
\cortext[cor2]{Principal corresponding author}
\fntext[fn1]{This is the first author footnote, but is common to third
  author as well.}
\fntext[fn2]{Another author footnote, this is a very long footnote and
  it should be a really long footnote. But this footnote is not yet
  sufficiently long enough to make two lines of footnote text.}

\nonumnote{This note has no numbers. In this work we demonstrate $a_b$
  the formation Y\_1 of a new type of polariton on the interface
  between a cuprous oxide slab and a polystyrene micro-sphere placed
  on the slab.
  }

\begin{abstract}
The cloud computing model enables the on-demand provisioning of computing resources, reducing manual management, increasing efficiency, and improving environmental impact. Software architects now play a strategic role in designing and deploying elasticity policies for automated resource management. However, creating policies that meet performance and cost objectives is complex. Existing approaches, often relying on formal models like Queueing Theory, require advanced skills and lack specific methods for representing elasticity within architectural models.
This paper introduces an architectural view type for modeling and simulating elasticity, supported by the Scaling Policy Definition (SPD) modeling language, a visual notation, and precise simulation semantics. The view type is integrated into the Palladio ecosystem, providing both conceptual and tool-based support.
We evaluate the approach through two single-case experiments and a user study. In the first experiment, simulations of elasticity policies demonstrate sufficient accuracy when compared to load tests, showing the utility of simulations for evaluating elasticity. The second experiment confirms feasibility for larger applications, though with increased simulation times.
The user study shows that participants completed 90\% of tasks, rated the usability at 71\%, and achieved an average score of 76\% in nearly half the allocated time. However, the empirical evidence suggests that modeling with this architectural view requires more time than modeling control flow, resource environments, or usage profiles, despite its benefits for elasticity policy design and evaluation.

\end{abstract}

\begin{graphicalabstract}
    \includegraphics[width=\textwidth]{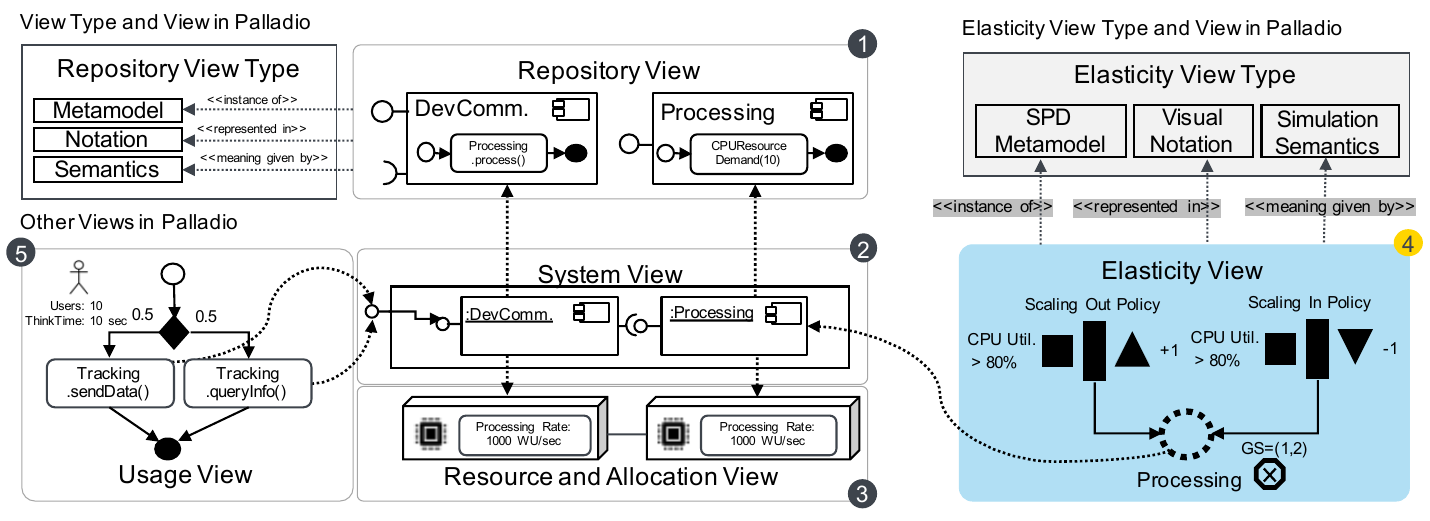}
\end{graphicalabstract}

\begin{highlights}

\item Proposes an architectural view type for elasticity modeling and simulation.
\item The view type includes the SPD metamodel, the visual notation, and the simulation semantics.
\item Evaluated based on predictive power and impact on productivity.
\item Accuracy and simulation efficiency measured via two experiments, modeling 16 different elasticity policies.
\item A user study shows modeling elasticity is more time-consuming than other aspects.

\end{highlights}


\maketitle

\section{Introduction}

Since the popularization of the cloud computing model over the last decade \cite{mell2011nist}, computing resources can be acquired and released in an automated fashion. This aligns with the overarching vision behind self-adapting software systems: to shift the human role from operational to strategic \cite{visionsas2010}. 
However, human operators design and configure the process responsible for the automated acquisition and release of resources. 
Therefore it is still possible to employ policies for automated management of resources---elasticity policies---that lead to high costs or bad performance. 
Decision makers like software architects are unaware of how the choices in the structural design space impact the functional behavior for a particular elasticity policy design.
For example, the architect can employ a policy available at the cloud provider that scales virtual machines or design one that is aware of the application components in the architecture.
Such structural choices influence the functional behavior of the policy.

From a software engineering perspective there are three alternatives available for the software engineer with concerns about the elasticity of the system: (a) employing and configuring ad-hoc exsiting proposals, (b) testing, and, (c) model-based analysis or model-based testing of policies. 
A vast number of concrete autoscaling and elasticity approaches are proposed both in literature~\cite{lorido2014review, chen2016survey, DBLP:journals/eaai/GariMPMG21} and practice from cloud providers~\cite{AWS:scaling:cooldowns} and popular orchestration frameworks such as Kubernetes or OpenStack~\cite{kubernetes_2022, OpenStackOrchestration}. 
The recent report of Flexera~\cite{flexera24} that surveys around 750 cloud adopters ranks cloud cost optimization as the top priority and challenge. Further, the report highlights that only 35\% of the surveyd companies have automated policies in place. 

The low adoption of automated policies may be attributed to mistrusting the behavior of the application in presence of such policies. Hence, one alternative to gain confidence that policies lead to satisfactory behavior in terms of qualities such as performance is to test. However, considering that cloud cost optimization is the main aim, testing autoscaling policies contributes with additional cloud cost. 

The third alternative of model-based testing or model-based performance engineering elasticity policies seems a viable option but has associated manpower costs for model creation, analysis time, and time for the interpretation of the results. 
A set of approaches require formal mathematical modeling skills such as Queuing Theory to create models \cite{1530877, DBLP:journals/simpra/VondraS17}.
Modeling approaches closer to the cloud computing domain and to the role of a software engineer, such as CloudSim~\cite{calheiros2011cloudsim} and Palladio~\cite{DBLP:journals/jss/BeckerKR09} do not support the modeling of elasticity concerns with first class modeling constructs. 
Hence, software architects must model the elastic behavior of cloud-native applications manually and with workarounds. This may lead to errors in model construction or long refining and debugging activities for the correct model construction.

To overcome the shortcomings of state-of-the-art approaches, this paper describes a design science based research that resulted in the Slingshot approach. 
We propose a dedicated architectural view type for the concern of elasticity modeling and simulation. 
The view type integrates to the Palladio approach for performance modeling and simulation of software architectures.
The view type is materialized by the following parts: 
\begin{itemize}
    \item The Scaling Policy Definition~(SPD) metamodel that defines the modeling concepts organized in 8 packages for modeling and simulating the elasticity of a software architecture. The metamodel features five core modeling constructs:
\texttt{ScalingPolicies} that apply to \texttt{TargetGroups}, which fire based on \texttt{ScalingTrigger}(s), and are constrained by a finite number of \texttt{Constraints}(s) that either apply to policies or to targets, and ultimately adjust the number of elements part of a target group by means of an \texttt{AdjustmentType}.
    \item A visual notation for the main modeling concepts.
    \item The simulation semantics that concisely defines the effect of simulating the elasiticty of the architecture based on the proposed modeling language by means of model to model transformations. 
\end{itemize}

The SPD metamodel is prpoosed as a result of following the design science methodology and identifying the factor that contributes to the inefficiency of evaluating elasticity alternatives. 
The visual notation is derived based on requirements for visual notations in software engineering~\cite{Moody2009} and shall aid the understandability of the modeling concepts. 
The last contribution is the simulation semantics of models instantiated in SPD which precisely defines what happens during simulation time when modeling elasticity in the proposed view type.
In this work, we describe all the three constituents of the proposed view type for elasticity modeling and simulation. 

To evaluate the contribution, we have conducted two representative single-case mechanism experiments to investigate the proposed approach's validity. 
We follow Wieringa's definition of a single-case mechanism experiment that defines it as a study in which the researcher intervenes and observes the impact of a new artifact or technology~\cite{wieringa2014design}. 
In the two cases, the approach allows the modeling of 16 different elasticity realizations. 
Compared to ground truth data, the best-case scenario exhibited a mean absolute prediction error of 23\%, while the worst-case scenario showed a 46\% error. 
These results, contrasted with the established literature criterion of a 30\% error margin for model-based performance engineering approaches~\cite{becker2013simulizar, Amiri2020}, indicate acceptable accuracy in the best-case scenario and require further improvements for the worst-case scenario. 
Additionally, utilizing the Slingshot approach for elasticity prediction significantly reduces computation time by two orders of magnitude compared to estimating actual values with load testing, underscoring the effectiveness of simulation-based elasticity policy design.
In a second case, the validation shows the feasibility of modeling elasticity policies for larger applications of representative size with the drawback of increased simulation time by one order of magnitude compared to the first.

Further, to validate the approach from an end-user perspective, we conducted a user study to collect 12 data points from a representative sample of master-level students who create and refine SPD models. 
The collected empirical evidence shows that novice software architects can model and refine elasticity policies in nearly half of the allocated time with an average correctness score of 76\%.
Participants rate the usability high, with an average system usability score of 71\%.
Modeling from scratch and refining existing elasticity policies constitute the two tasks of the experiment that represent the intended use cases of the approach.
When comparing the empirical evidence with the existing empirical study of associated modeling effort in Palladio~\cite{Martens2008}, there is an indication that modeling elasticity through the proposed architectural view yields lower throughput 
than modeling the control flow, the resource environment and the usage profile.
An explanation that requires further validation is that the specification of dynamic behavior is harder to comprehend. 
Results indicate the utility of the proposed architectural view in terms of reduced modeling effort, reduced analysis time compared to load tests for autoscaling policies, and positive perception by end-users.

The structure of this paper is as follows. 
The next section introduces a running example for explaining the contribution. 
The running example describes also the first experimental case used for the evaluation.
The follow-up section provides foundational background by elaborating the Palladio approach, elasticity concerns, and how elasticity concerns are modeled and simulated within Palladio.
In the foundation section we additionally elaborate the terminology followed in this work.  
The methodology section describes how we instantiated the design science research method for understanding the problem, defining the design goals, approaching a solution and designing the evaluation for the proposed artefacts. 
The contribution is elaborated by describing the abstract syntax, the visual notation, and the simulation semantics for resulting models created through the proposed view type.  
After presenting the results and discussion of the contribution, the final section covers related works with highlights on the difference to the proposed approach and justifications for the research gap which we address with the contribution. 

All artifacts, including code and experimental data are open and accessible~\cite{FlorimentKlinaku2024}.
The proposed approach is technically realised in the Eclipse Modeling Framework~\cite{Steinberg2008}. 
The technical realization of the language and the simulator that is used for the evaluation part of this work are part of the Palladio GitHub Organisation\footnote{\url{https://github.com/PalladioSimulator?q=Slingshot}}.

\begin{figure}
    \centering
    \includegraphics[width=0.95\columnwidth]{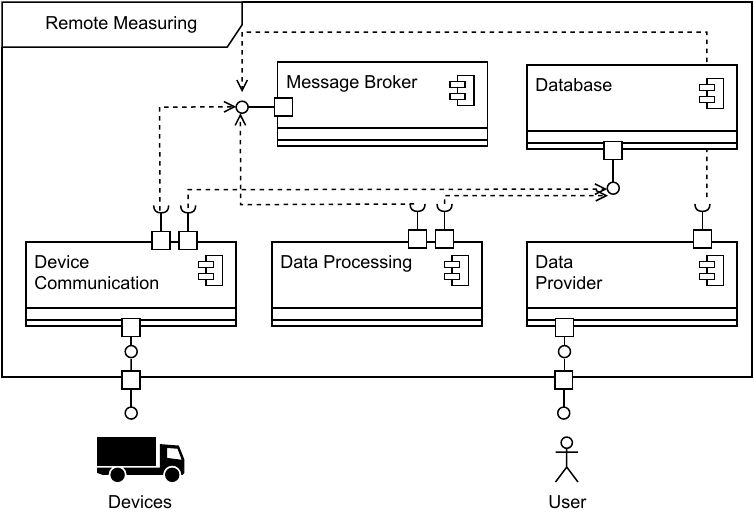}
    \caption{Architecture of the Remote Measuring Use Case.}
    \label{fig:remote-use-case-system}
\end{figure}

\newpage
\section{Running Example}
As a running example, to motivate the topic and elaborate the contribution of this work, we use the Remote Measuring Use Case (RMUC) application. 
The application has been designed and developed in the reseach project MOSAIC - Modelling, Simulation and Design of Self-Adaptive IoT  Systems in the Cloud and is publicly available~\cite{Research2023}.

The cloud-based application allows users to manage measurement campaigns from remote devices, e.g., vehicles, vessels, or other devices connected to the Internet (see \Cref{fig:remote-use-case-system}).  
The system offers two interfaces. 
One interface allows remote devices to push data recorded during a measurement campaign. 
The system offers a second interface to allow technicians to query measurement campaigns. 

To realize these two main functionalities, the system is decomposed in three main business components and two technical components to handle communication and persistance of the date. \Cref{fig:remote-use-case-system} shows the decomposition of the application and their intereaction. 
One or more devices send data from their bus to the Device Communication service. 
The Device Communication service tracks the readiness of the device, caches the data, and stores the raw data in the database. 
On the other hand, when realizing the second use case,  end users (technicians) can query a certain measurement campaign. 
In that case, the Data Provider service communicates with the Data Processing service to retrieve the queried data. It converts the data into a presentable and readable format, different from the format Database uses to store measurement campaigns.

\begin{figure}[ht]
    \centering
    \includegraphics[width=0.65\columnwidth]{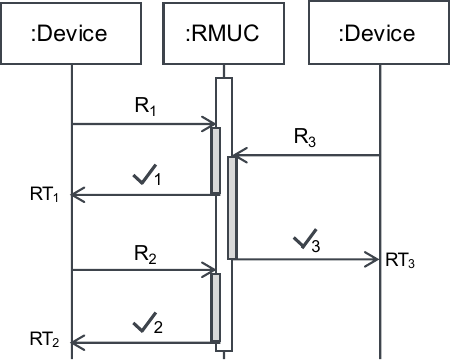}
    \caption{Send Data Elasticity Scenario.}
    \label{fig:remote-use-case-scenario}
\end{figure}

\subsection{Elasticity Scenario}
\label{sec:scenario}
Diverse usage patterns characterize the case. The number of devices is unpredictable, and campaigns may vary significantly in terms of number of recorded vehicle signals and their recording frequency.
Furthermore, since the case is planned to be deployed on a public cloud platform, there is inherent uncertainty driven by workload placement, sharing the resources with others, and other factors.
Despite the diversity and uncertainty in the environment, stakeholders are interested in providing a specific level of performance.

Through the scenario in \Cref{fig:remote-use-case-scenario} we illustrate the elasticity objectives behind the case. 
In the scenario two devices send data to the RMUC application. Upon receiving the request and processing it, the system produces an acknowledgment. 
After receiving the acknowledgment, the logged time $RT_{1}$ denotes the response time for a device sending data to the system. 
The average response time for devices sending data denotes the average time of all such requests that occur during a measurement campaign. 
stakeholders  It is the interest of.
The elasticity concerns of stakeholders in this case are to keep the performance at a desired level in presence of the aformenetioned uncertainties e.g., diverse usage patterns. 
Hence, a the specific level of performance is formulated as follow: \enquote{the RMUC application shall process 95\% of the devices sending data during a campaign in less than 1 second.}
The fulfilment degree of such an objective is influenced by relevant architectural design decisions~\cite{Jansen2005}.
Relevant decisions include aspects like how the system is decomposed, how do components interact, how they are allocated on the provisioned resources and how resources are managed during a measurement campaign. 
Such decisions yield the software architecture of the system. 
One set of decisions of relevance for this paper are decisions about how resources are autonomously managed through elasticity policies.  
We name such concerns as elasticity concerns.

\subsection{Elasticity Concerns}

Architectural elasticity  concerns for the presented case include the following: 
(a)~determining the impact of employing autoscaling policies on performance and cost objectives, 
(b)~determining the impact of colocating business components and autoscaling the hosting node, 
(c)~determining the impact of autoscaling some of the business components but not all, 
(d)~determining what architectural information shall be used for autoscaling, 
(e)~determining the impact of the employed autoscaling policies under different workload levels, 
(f)~determining the impact of constraining autoscaling policies because of budget, 
(g)~determining the impact of using different configurations of the autoscaling process. 
Making decisions for the elasticity of the system is influenced by prior architectural decisions such as having a certain decomposition of the system or following a certain deployment pattern.

\clearpage
\section{Foundations}
The contribution proposes an architectural view type concerning elasticity for quality of service predictions at design time. Hence, in this section we elaborate two foundational topics: the Palladio approach for performance engineering based on architectural models and the prerequisital understanding for elasticity concerns. In the first part, we contextualise the terms view type, metamodel and model transformations that are neccessary to understand the solution approach.

\subsection{The Palladio Approach}
\label{sec:palladio}

\begin{figure*}
    \centering
    \includegraphics[width=0.6\textwidth]{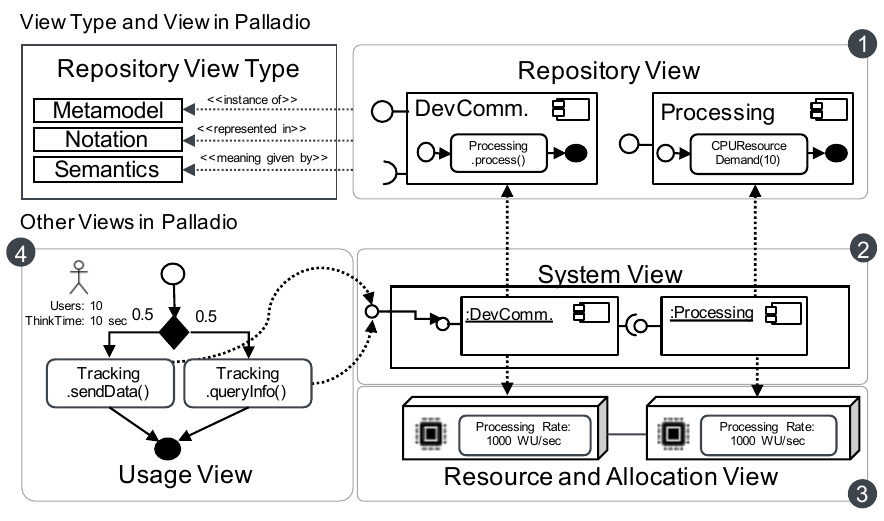}
    \caption{Modeling Views in Palladio for the Running Example.}
    \label{fig:palladio-overview}
\end{figure*}

The contribution builds upon the Palladio approach~\cite{reussner2016modeling} which is a state of the art approach in the area of component-based software performance engineering where the quality of a software is predicted based on architectural models in the early design stage~\cite{williams1998performance}.
In the Palladio approach, the modeling of a software system is distributed in concern specific views belonging to different viewpoints. 
The complete model is simulated to predict various quality of service attributes such as response times, throughput or number of resources used and their utilization.  
The Palladio approach is organized through three viewpoints: structural, behavioral, and allocation. 
Each viewpoint consists of multiple view types to allow the modeling of the system and the representation of specific concerns. 
The viewpoints in Palladio align with the broader definition of architectural viewpoints in the ISO 42010 standard~\cite{9938446} where a viewpoint is defined as \enquote{frame of reference for the concerns determined by the architect as relevant to the purpose of the architecture description}.
A view type is defined by a metamodel that precisely defines the concepts, rules and constraints for creating view instances~\cite{baier2008a}, a mapping of the main concepts to visual symbols (graphical syntax), and the simulation semantics of the resulting models.
In this work, we contribute the elasticity view type that frames elasiticity concerns for simulation purposes. 
The elasticity view type contributes to the viewpoint of self-adaptivity or reflection of the software system in Palladio. 
View types in the self-adaptivity viewpoint influence and impact view types in the three original viewpoints.




The upper part of \Cref{fig:palladio-overview} depicts the relation of one of the existing views---the repository view---and its view type. For each view there exist a well-defined metamodel, graphical notation and semantics. 
The semantics is defined by the interpretation of the models in simulators or by transformation of the models into analysis models such as layered queuing networks. 
The lower part of the figure shows additional views that exist in Palladio. We describe briefly the concerns in each view. 

The repository view enables the modeling of software components, their provided and required interfaces and the internal behavior in terms of resource demands. 
In the \Cref{fig:palladio-overview}, the resulting repository view of the example model~(\ding{202}) contains two components named Device Communication (DevComm.) and Processing. 
A software component describes the behavior in terms of performance with context aspects parameterized. 
The most important constituents of a component in Palladio are the Service Effect Specifications (SEFFs). 
A SEFF is an ordered sequence of actions of different types.
The different types determine resource demands and execution control flow for a simulated user.
For example, an internal action type may define the resource demand for a particular resource type, e.g., 10~CPU work units for a certain computation.


There exist other views in Palladio that have their own metamodel, notation and semantics. 
For example, in the system view, the modeled components in the previous step are instantiated and linked to form a system that provides an interface~(\ding{203}). 
In the resource and allocation view, the instantiated components are allocated onto two resource containers that have the same processing capability of 1000 work units per second~(\ding{204}). 
Finally, in the usage view, the example models that the system is exercised by 10 users with a think time of 10 seconds and that in 50\% of the cases data is sent to the system and in the other 50\% the data is queried~(\ding{205}).

\newpage
\subsection{Elasticity Concerns}

\begin{figure*}[t]
	\centering
 	\includegraphics[width=0.8\textwidth]{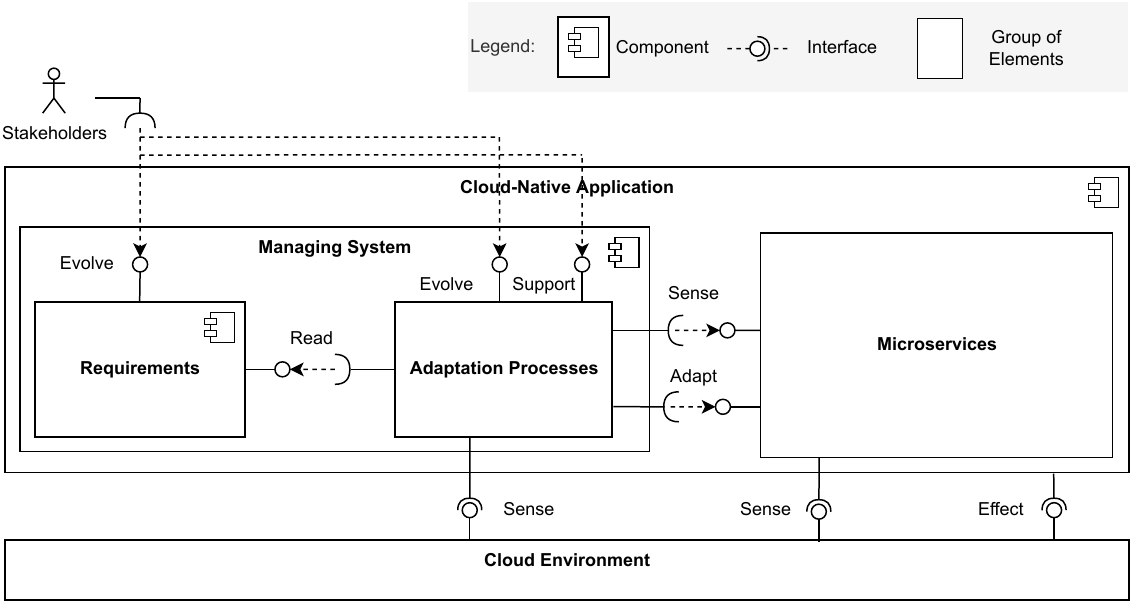}
	\vspace{-10pt}
	\caption{Conceptual architectural model for a cloud-native application based on Weyns conceptual model~\cite{weyns2020introduction}.}
	\label{fig:conceptual-model-cna}
\end{figure*}

The proposed view type adresses elasticity concerns. Elasticity is an esential non-functional requirement for applications that utilize the cloud computing service model known as cloud-native applications. The adjective \enquote{cloud-native} describes the highest maturity level of an application deployed on a cloud platform, where the main property entails the ability to exploit the elastic and scalable cloud platform to adjust capacity on-demand~\cite{kratzke2018brief, ashtikar2014open}. 
One main property for cloud-native applications is elasticity. Several definitions exist for elasticity in the context of cloud computing. We follow the definition below from Herbst et al.~\cite{DBLP:conf/icac/HerbstKR13} that define elasticity as \enquote{the degree to which a system is able to adapt to workload changes by provisioning and de-provisioning resources in an autonomic manner, such that at each point in time, the available resources match the current demand as closely as possible}~
\cite{DBLP:conf/icac/HerbstKR13}. 

Following the definition above, elasticity is a property of a cloud-native application engineered to automatically provision and release resources with the objective to fulfill the two following relaxed requirements to a certain degree: 1. $R_{QoS}$: under unexpected load variations, the elasticity mechanism shall maintain the QoS level as close as possible to a desired target, and, 2. $R_{Cost.}$: under unexpected load variations, the elasticity mechanism shall provision as few as possible cloud resources incurring costs.

Software systems that adapt their architecture at runtime are known in the literature as self-adaptive software systems~\cite{weyns2020introduction}. 
Self-adaptive software systems have emerged as a class of software systems that incorporate feedback loops to reconfigure structure or parameters at runtime, given the environmental changes.
Cloud-native applications can be seen as a subset of self-adaptive software systems that utilize the elastic platform to provision resources based on service demands. 

Considering cloud-native applications a subset of self-adaptive software systems, \Cref{fig:conceptual-model-cna} shows a conceptual architectural model for cloud-native applications based on the conceptual model by Weyn~\cite{weyns2020introduction}.
In this model, adaptation processes sense and adapt microservices that all together realize a certain business capability. 
An adaptation process is a self-contained deployment unit that realizes a feedback loop that senses the environment and changes the number of microservices running according to given requirements. 
One prominent example of an adaptation process is the Horizontal Pod Autoscaler (HPA) from the well-known container orchestration engine Kubernetes~\cite{kubernetes_2022}.
The HPA reads requirements from a YAML configuration file that usually specifies thresholds on metrics sensed from the environment, such as CPU utilization.
The autoscaler adapts the number of containers when the threshold is exceeded. Upon adaptation, containers are created and scheduled on available nodes.
The stakeholders, usually DevOps engineers, can evolve the requirements by changing the YAML configuration, evolve the adaptation process by altering its internal workings, and perform support by monitoring its behavior and acting accordingly.

Elasticity concerns include concerns of determining the QoS impact of certain design decisions for the autoscaling process.
Based on the conceptual model in \Cref{fig:conceptual-model-cna}, such decisions stem from the different parts of the architecture: what information the adaptation processes sense from the environment and from the managed components, what type of analysis is performed on that information, what type of adaptations are enacted by the adaptation process, what is the impact of varying the requirements to the adaptation process (e.g., reacting on different thresholds), what has to be monitored and what would be the impact of evolving the requirements or the internals of an adaptation process.

\clearpage
\subsection{Answering Elasticity Concerns with Palladio}

The Palladio approach and the corresponding view types support the modeling and analysis of static system configurations i.e., the allocation of components on resources remains fixed during a simulation run. 
For analyzing self-adaptivity concerns including elasticity, the modeled concerns across the different view types have to be modified or transformed during the simulation. 
\begin{figure*}
    \centering
    \includegraphics[width=0.7\linewidth]{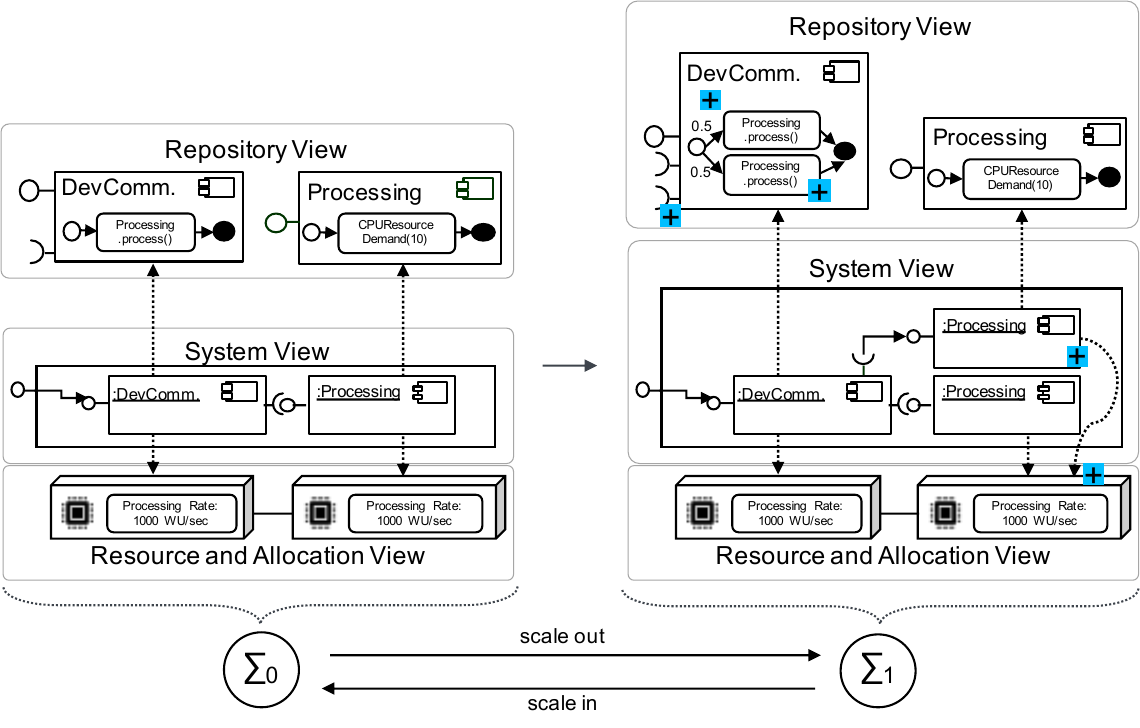}
    \caption{Simulation of an  Elasticity Policy based on Palladio.}
    \label{fig:lts}
\end{figure*}
The two definitions below are taken from Becker~\cite{Becker2017} and define the reconfiguration space and a single state of the modeled self-adaptive system state and the space explored during a simulation run of a model that changes its structure during simulation.

\begin{definition}[Self-Adaptive System Reconfiguration Space]
\label{def:lts-sas}
A self-adaptive system reconfiguration space is a labeled transition system $\Gamma$ = ($\Sigma$, $\rightarrow$, $\Sigma_0$), in which
\begin{itemize}
	\item $\Sigma$ is the set of all self-adaptive system states;
	\item $\rightarrow$ is the labeled transition function;
	\item $\Sigma_0$ is the initial self-adaptive system state.
\end{itemize} 
\end{definition}

\begin{definition}[Self-Adaptive System State]
\label{def:lts-sas-state}
 A self-adaptive system state  $\Sigma_t \in \Sigma$ is defined by the tuple ($a$, $s$, $m$, $t$), in which
	\begin{itemize}
		\item $a$ is an element from all possible architecture configurations, $a \in A$
		\item $s$ is an element from all possible context scenarios, $s \in S$
		\item $m$ is an element of the set of run-time measurements, $m \in M(a, s)$;
		\item $t$ is a point in time, $t \in T$.
	\end{itemize}
\end{definition}

The labeled transition function $\rightarrow$ advances the simulation by applying distinct operations to the initial state $\Sigma_0$. One operation maps the current state of a property to a numerical metric, such as the mean response time of an operation at a specific moment. This metric serves as a runtime measurement to guide system adjustments. Another operation uses the current system configuration and a set of conditions on runtime measurements to determine a new configuration. For instance, a condition may check whether the mean response time exceeds a predefined threshold. This new configuration is generated by applying transformation rules to the model. Finally, the labeled transition function can also account for changes in the system's usage context over time, capturing the evolving scenarios that influence system behavior.

\Cref{fig:lts} depicts a simulation in which the architecture may transition to a state in which there exist two replicas of the Processing component and may scale back in to the initial state based on a different condition. 
The modeling of such an elasticity policy in existing works may be accomplished in three different forms. 
In one of the forms, the elasticity is modeled by relying on generic transformation languages such as QVTo or Henshin in which the modeler has to correctly transform the models to reflect a scaled configuration~\cite{becker2013simulizar}. In a second form, the modeling of elasticity concerns is achieved by  annotating elements in existing views with properties that define the scaling conditions and the effects of scaling in terms of adaptations~\cite{DBLP:phd/dnb/Lehrig18}. The third form involves the reuse of existing views and elements to model the performance impact of adaptations and model the scaling effects as model transformations~\cite{Stier2017}.
The evaluation of existing approaches motivates the contribution for improving the productivity when evaluating elasticity concerns.

\clearpage
\section{Methodology}

The proposed view type for elasticity modeling and simulation stems from following the design science research method~\cite{wieringa2014design}.
Wieringa defines design science as the investigation and improvement of artefacts in a predefined context~\cite{wieringa2014design}.

\subsection{Problem Understanding and Design Goals} 
We approach the understanding of the problem with two goals: (1) understanding the peculiarities for cloud-native applications, as well as understanding how responsible roles (e.g., DevOps engineers or software architects) decide on elasticity policies for cloud-native applications, and (2) understanding how state-of-the-art methods, such as Palladio, support the decision making for elasticity policies.

To understand the problem, we initially conduct an observational case study to evaluate Palladio and its support in answering design concerns with respect to performance, cost-efficiency and elasticity~\cite{Klinaku2021}.
Scenarios in relation to elasticity have been evaluated as infeasible and requiring workarounds. 
When modeling elasticity policies for the scoped case~\cite{10.1145/3344948.3344961}, it is manageable to evaluate various elasticity policies, but we mark the reliance on general purpose transformation languagues such as QVTo~\cite{qvto} as the main factor that yields an inefficient modeling, simulation and feedback process.
Relying on general-purpose languages~(GPL) affects all aspects of elasticity policy modeling.
The modeler's expertise directly influences the effort in modeling elasticity policies, i.e., how much time is spent, as well as, the resulting quality of the model, i.e., whether it contains errors. 
In addition, relying on general purpose languages means that the simulation has to execute the transformations, but the modeler's code may be inefficient and thus worsen simulation time. 
The last aspect is the missing feedback link. It is hard to establish a link between the simulation output and the artifact as code to allow the modeler to refine the elasticity alternative. 
For example, feedback related to when the policy has triggered, under what conditions, how many instances were added have to be obtained either via debugging in the best case or are not retrieval at all.  

In addition to investigating the problem in the context, we conduct an in-depth literature review to show state-of-the-art approaches related to the elasticity modeling at design-time.  
Using the literature review, we highlight the gap and elicit the design goals and the requirements. 
We highlight the outcomes of the literature review after discussing the contribution of the paper.
The literature review also aided as of the resources for eliciting the main concept of the proposed view type for elasticity.  

The higher level design goal is the efficient design of elasiticy policies. We decompose this goal into three sub-goals: modeling efficiency, simulation efficiency, and feedback efficiency.
Related to the derived goals, existing works hamper all the three aspects by mainly requiring from architects to rely on general purpose languages to model the elasticity of the system.

\subsection{Solution Design}

We construct the metamodel that consists of the main concepts available in the proposed view type through an iterative process as defined in~\cite{DBLP:journals/spe/StrembeckZ09} for the systmatic development of domain specific languages. 
The first step of the process consists of identifying elements in existing platforms. 
Three data sources influenced the elicitation of the main modeling concepts and their relationships. 
We extract domain concepts from existing languages from the Palladio context 
as well as from domain-specific languages for elasticity. 
Additionally, we collect concepts present in public cloud providers for configuring the elasticity of applications. 
The third data source that influenced the main design decisions are the works in the area of policy-driven management of distributed systems~\cite{DBLP:conf/policy/DamianouDLS01, DBLP:journals/jnsm/Sloman94} that propose the decoupling of policies from managers that realize the policies. 



The graphical syntax results from a graphical editor constructed in a course of a bachelor thesis~\cite{Summerer2022} where the guidelines by Moody~\cite{Moody2009} for constructing visual notations in software engineering have been followed. 
Nine participants of diverse background have rated the decisions on whether the symbols are appropriate to represent the modeling concept through yes/no questions.
The symbol for the following modeling concepts have been evaluated: scaling policy, adjustment type, direction-dependent adjustment type, scaling trigger, constraints and scaling targets.  

The simulation semantics gives precise meaning to the modeled elasticity policies or alternatives.
Technically, in the concept used for the evaluation, the simulation semantics have been implemented in QVTo and integrated into the Slingshot simulator. 
The simulation semantics defines precisely and completely the reconfiguration space~(c.f., \Cref{def:lts-sas}) given a Palladio model in conjuction with an instance of the metamodel for the contributed elasticity view type. 

\subsection{Evaluation Design}
The following research question confines the evaluation of the contribution.
\begin{enumerate}[label=\subscript{R}{{\arabic*}}]
	\item[$RQ$] What is the \textit{efficiency} of the Slingshot approach for evaluating elasticity alternatives at design time?
\end{enumerate}
The efficiency aspect can be attributed to the three components of the design goals: modeling efficiency, simulation efficiency and feedback efficiency. 
For evaluating modeling and simulation efficiency, we conduct two single-case experiments. 
In the first, we infer the prediction accuracy and associated effort for evaluating elasticity alternatives in comparison to load testing which is a common employed practice. We use the second experimental case to evaluate the simulation efficiency for simulating the elasticity of larger case.
For evaluating modeling and feedback efficiency, we design and conduct a user experiment where representative users for the population of software architects were tasked with modeling and refinement of elasticity alternatives in the proposed view. 

.
\clearpage
\section{The Elasticity Architectural View Type---The Slingshot Approach}

\begin{figure*}[t]
    \centering
    \includegraphics[width=\textwidth]{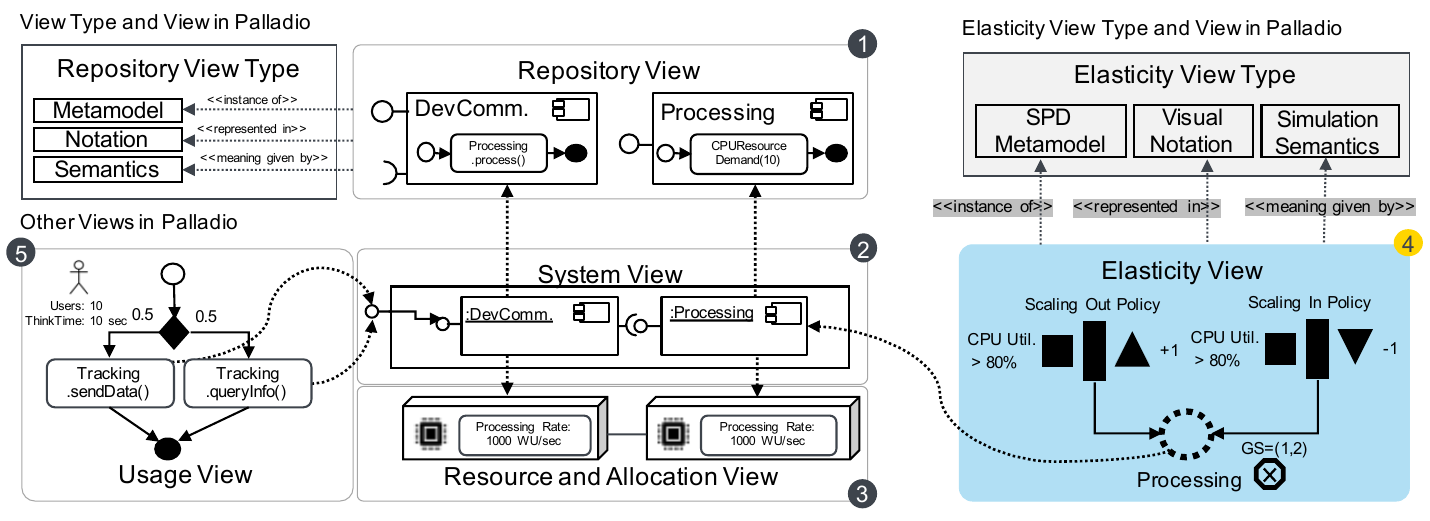}
    \caption{Overview of the Slingshot approach for Elasticity Modeling and Simulation.}
    \label{fig:approach-overview}
\end{figure*}

The Slingshot approach extends the Palladio approach with a new view type for modeling the elasticity of a software architecture, i.e., with scaling policy definitions that define what parts of the architecture scale when. 
In the Palladio development process, the Palladio Component Model instance is obtained when several roles model the system under analysis from various viewpoints.
\Cref{fig:approach-overview} depicts the resulting models in the different views. 
The models created during modeling time define the $Palladio_{init}$ model. 
In the $Palladio_{init}$, the instantiated components in the Assembly and their allocation in the Allocation model are fixed and remain unchanged during a simulation run. 
Through the new architectural view type for elasticity, the self-adaptive system architect creates scaling policies, which are interpreted during simulation time, and as an effect, the initial Palladio models are transformed during simulation time.

Deciding what part of the architecture is configured to be scaled and what parts of the architecture remain at their initial configuration defines one important design decision that we name as \textit{identifying the Slingshot}. 
We name the approach as the Slingshot approach by following the analogy of how the engineering of an actual slingshot takes into consideration the physical elasticity properties of the rubber strips and the rigidity of the Y-shaped frame.
For example, the self-adaptive system architect may define policies only for one of the instantiated components and only that part is impacted through the simulation while other remain at their initial configuration.
This is depicted on the right part of \Cref{fig:approach-overview} where only the component \texttt{Processing} scales due to the applied policies while the component \texttt{DeviceCommunication} remains at the initial size.
Technically, to transform the models correctly according to the defined semantics, the self-adaptive system architect creates a semantic configuration that configures elements from the $Palladio_{init}$ to be scalable and controllable via elasticity policies that are defined as a subsequent step through the proposed elasticity view type. 

The next subsections describe the following aspects of the Slingshot approach: 
In the \textit{SPD Metamodel} section, we describe the requirements we want to fulfill and the resulting metamodel that contains concepts and relations that a software architect could use in the proposed elasticity view. 
The abstract syntax features five core modeling constructs: scaling policies, scaling triggers, constraints, target groups and adjustment types.
In the \textit{Visual Notation} section, we describe the designed visual notation for the modeling constructs and discuss an example in such a notation. 
As shown in \Cref{fig:approach-overview}, in each view, there is a visual notation to help the corresponding role understand, comprehend, and easily create model instances.
In the \textit{Simulation Semantics} section, we describe on a high level the transformation rules that transform the initial Palladio models in the presence of SPD models. We put special focus on differentiating between a top-down transformation semantic in which scaling policies are defined for the services and a bottom-up transformation semantic in which scaling policies are defined for the infrastructure. 

The Slingshot simulator is a discrete-event simulator that is able to interpret Palladio models enriched with elasticity models created in the proposed architectural view type~\cite{Klinaku2023}. 
There exist two world-views for discrete event simulations: process oriented and event oriented~\cite{Carson}. 
In the process oriented world view, the dynamic behavior is modeled as a sequence of time-ordered activities and delays whereas in the event-oriented world view, event routines execute in zero simulated time and advance the state~\cite{Carson}. 
To offer extensibility and the simulatability of self-adaptive models, the Slingshot simulator follows an event-oriented simulation worldview.
In addition, it relies on an event-driven architecture in which simulation behaviors communicate through events to notify state changes.  
All the models in the \Cref{fig:approach-overview} are an input to the simulator. 
The simulation behavior that interprets SPD models relies on the simulation semantics presented in this work. 
The internal workings of the simulator itself are out of the paper's scope. 

\clearpage
\subsection{SPD Metamodel}
\begin{figure*}
    \centering
    \includegraphics[width=\textwidth]{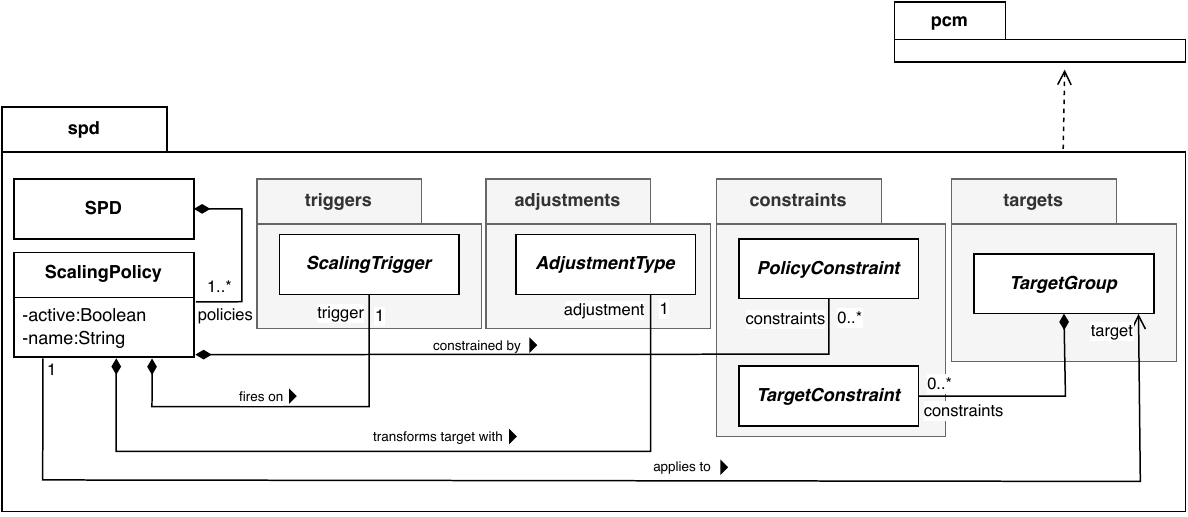}
    \caption{The core classes and relationships of the Scaling Policy Definition language.}
    \label{fig:spd-core}
\end{figure*}

The core classes and their relationships entail several essential design decisions.
The first class in \Cref{fig:spd-core} is the root element named \texttt{SPD} that stands for Scaling Policy Definition.
An \texttt{SPD} is the containment of at least one \texttt{ScalingPolicy}. Both classes are part of the \textit{spd} root package.
The containment relationship allows the architect to create and analyze several scaling policies in composition. For example, it will enable the analysis of one scaling policy responsible for scaling out the application and another for scaling in the application.
In addition, the \texttt{ScalingPolicy} has an attribute named \texttt{active}, which allows architects to toggle policies on and off.

A single \texttt{SPD} defines the unit of analysis, i.e., the \texttt{SPD} is an input to the simulation, and the results yield the system's performance or other quality dimensions under a single \texttt{SPD}.
The language allows the definition of multiple \texttt{ScalingPolicies} in one such unit of analysis.
This way, self-adaptive system architects can define multiple policies that govern the adaptation of the application under investigation.

Each policy has its trigger that defines when a scaling decision occurs---aligning with the requirement of being concise regarding what triggers scaling.
Each \texttt{ScalingPolicy} encapsulates the information necessary to transform the model to a different scaled state.
That information is distributed over four core classes in four sub-packages to separate concerns and facilitate extensibility.
These classes are the \texttt{TargetGroup} (\textit{targets}), the \texttt{AdjustmentType} (\textit{adjustments}), the \texttt{ScalingTrigger} (\textit{triggers}), and the \texttt{Constraint} (\textit{constraints}).
The \texttt{ScalingPolicy} \textit{adjusts} the model with precisely one \texttt{AdjustmentType} whenever exactly one \texttt{ScalingTrigger} \textit{fires}.

In addition, the \texttt{ScalingPolicy} is \textit{constraint by} a collection of \texttt{Constraint}(s) and \textit{applies to} a single \texttt{TargetGroup} through referencing. 
A \texttt{TargetGroup} can contain several \texttt{Constraint}(s) that customize the target group's behavior irrespective of the independent policies.

\paragraph{Target Groups.}

In the SPD modeling language, target groups represented by the abstract class \texttt{TargetGro\-up} define what is being scaled or adjusted regarding capacity. 
It is the structure on which the modeler can apply various \texttt{ScalingPolicies}. 

A \texttt{TargetGroup} is an abstraction that logically defines a set of elements that supports two operations: adding a finite number of elements to the set and removing a finite number of elements from the set.
An interpretation or simulation is responsible for giving precise meaning to how the addition and removal of elements to the set is mapped to the actual technology or model.

We elicit three concrete refinements: the \texttt{ElasticInfra\-structure}, the \texttt{ServiceGroup}, and the \texttt{CompetingConsumersGroup}. 
The refinements allow the modeler to specify scaling policies that are aware of the application and policies that are agnostic of the structure. 
With the \texttt{ElasticInfrastructure} construct, we support the definition of scaling policies at the infrastructure level, agnostic to the hosted components.  
In the case of defining policies that are aware of the application architecture, we identify patterns that form management domains with specific and diverse interactions. Two such structures are already part of the metamodel: the \texttt{ServiceGroup} and the \texttt{CompetingConsumersGroup}. 

Similarly, Fehling et al.~\cite{CloudComputingPatterns2014} distinguish three patterns of elasticity realizations which also assume different structures to be managed.
The \enquote{Elasticity Manager} pattern is a realization based on utilizing the cloud resources on which the application is hosted.  
The \enquote{Elastic Load Balancer} pattern is a realization based on the number of synchronous accesses. 
It assumes the existence of a load balancer that distributes loads across a set of replicas. 
The \enquote{Elastic Queue} pattern is a realization that adjusts the number of required application components based on the number of queue messages. 

Contrary to Fehling et al.~\cite{CloudComputingPatterns2014}, the three target group refinements do not presume the triggering that causes an adjustment.  
Instead, they align to the patterns as structures that can be automatically managed at runtime. 
More such structures could be created by inheriting from the parent \texttt{TargetGroup} class without impacting definitions of scaling policies for previous structures. 

\begin{figure*}
    \centering
    \includegraphics[width=\textwidth]{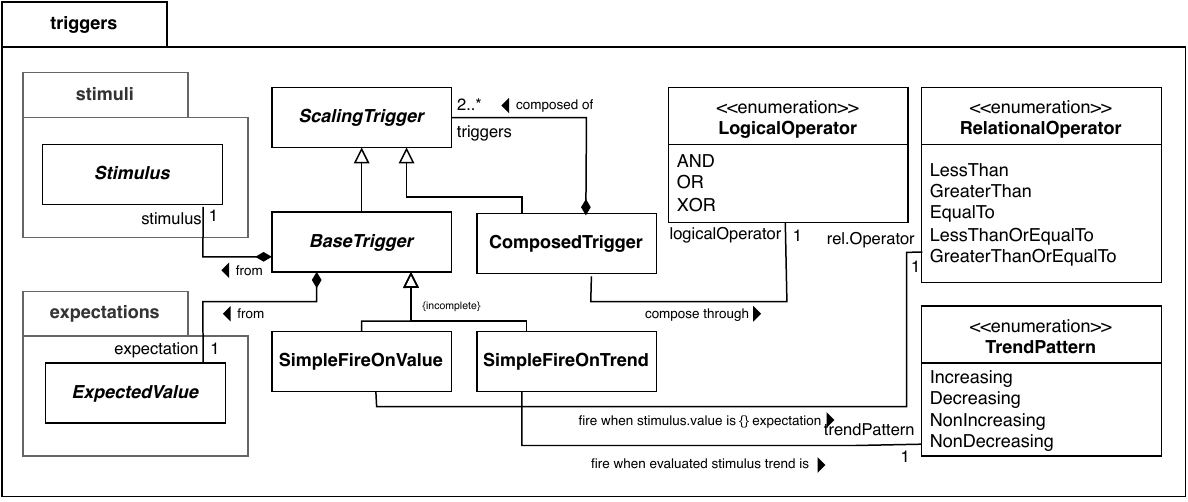}
    \caption{The core classes for defining triggers in SPD.}
    \label{fig:spd-core-triggers}
\end{figure*}
\paragraph{Triggers.}
The main design decision concerning triggering is to merge the definition of the sensed or monitored information in the environment, and the analysis into one construct called the \texttt{ScalingTrigger}.
The state of the target group, the simulation time, and the duration of requests waiting entail the main categories of information in the environment that can be sensed.
The dimension of what can be monitored in the environment is orthogonal to the type of analysis a scaling mechanism could conduct. 
\Cref{fig:spd-core-triggers} depicts the class hierarchy of the triggers in SPD. 
The primary driver requirement is to find a declarative, natural, and easy-to-understand way of modeling the condition \textit{when} a policy adjusts the target group. 
The core classes behind the trigger metamodel entail the \texttt{ScalingTrigger}, the \texttt{Stimulus}, and the \texttt{ExpectedValue}. 
The definition of a trigger and its constituent, a stimulus, and an expectation should allow the architect to define naturally, declaratively, and in an easy-to-understand way when an adjustment to the target group is made. 
The triggering concept has entangled two orthogonal dimensions: the information of what can be sensed in the simulation model---the stimulus---and the transformation of that information into a trigger. 
Hence, we create two different class hierarchies and bridge them in a \texttt{BaseTrigger} that contains the \texttt{Stimulus} and the \texttt{ExpectedValue}. Concrete refinements of the \texttt{BaseTrigger} abstract class determine how the processing of the stimulus and the expected value constitutes a trigger. Two refinements are already present, the \texttt{SimpleFireOnValue} that additionally requires a relational operator to allow the architect to connect the stimulus and the expected value through it. The \texttt{SimpleFireOnTrend} refinement on the other side, relies on a trend pattern to allow triggering when the stimulus is following a certain trend, e.g., increasing. 
To be succinct, we omit the class hierarchies for the \texttt{Stimulus} and \texttt{ExpectedValue}. For stimuli, there are concrete classes present to allow the triggering on the aforementioned aspects of the simulation model. For example, a \texttt{CPUUtilization} for the utilization state of CPU for a target group, a \texttt{QueueLength} for the queue length of requests waiting to be processed by a target group.
Similarly, for the expectations, there exist concrete classes such as \texttt{ExpectedPercentage} to define a double value that defines the expected percentage double value, which when bridged with a \texttt{CPUUtilization} stimulus in a \texttt{SimpleFireOnValue} with a given \texttt{RelationalOperator} determines the complete information required to trigger a scaling adjustment of a target group. 

\paragraph{Adjustments.}
The \texttt{AdjustmentType} defines the connecting point between the triggers and the target for which the policy applies.
The \texttt{AdjustmentType} defines in a declarative way how the input argument for the interface is computed. 

We define three initial refinements of the \texttt{AdjustmentType} class based on literature and the offering of cloud providers.
Upon firing a trigger through the \texttt{AbsoluteAdjustment} subclass, we capture the ability to scale to a specific capacity.
If the architect needs to scale relatively to the current capacity, we define the concept of \texttt{RelativeAdjustment} that allows the input of a growth percentage value. The \texttt{StepAdjustment} shall allow the architect to express a value which, upon the firing of a trigger, is added or removed from the current set of elements in the target group. 

The parameters of \texttt{RelativeAdjustment} and \texttt{StepAdjust\-ment} can be negative integers to define a decrease in capacity. 
The parameter of the \texttt{AbsoluteAdjust\-ment} is constrained to a positive integer.
However, depending on the current capacity, it can also express scaling in decisions, i.e., an elastic infrastructure with 10 resource containers scales in when an \texttt{AbsoluteAdjustment} with a goal value of 5 is made.

For an \texttt{AbsoluteAdjustment}, given a positive integer \(n~>=~1\) for the current number of elements of a target group, a positive integer value \(a\) for the \texttt{goalValue} then the new number of replicas $n'$ is equal to $a$. 

For a \texttt{RelativeAdjustment}, given a positive integer \(n >= 1\) denoting the current number of elements for a target group, an integer value \(p\) for the \texttt{percentageGrowth} and a value \(m\) for \texttt{minAdjustment} then the new number of replicas is as follow: 
\begin{equation*}
\label{eq:relativeadjust}
\begin{cases}
	n' = n + max(\lceil \frac{n*p}{100}\rceil,m) \text{, if } p>0, m>0 \\
	n' =  max(1, n - max(\lfloor\frac{n*|p|}{100}\rfloor,|m|) ) \text{, if } p<0, m<0 \\
\end{cases}
\end{equation*}

For a \texttt{StepAdjustment}, given a positive integer \(n >= 1\) denoting the current number of elements for a target group, an integer value \(s\) for the \texttt{stepValue} then the new number of replicas is as follow: $n' = max(1, n + s)$

\paragraph{Constraints.}
Based on literature (e.g., \cite{DBLP:journals/jnsm/Sloman94,DBLP:conf/policy/DamianouDLS01}) and the continuous refinements we distinguish three orthogonal dimensions for modeling constraints: the \textit{type}, the \textit{scope} and the \textit{behaviour}. 

The \textit{type} characterizes the constraint, whether the reconfiguration space is constrained based on time (temporal) or the state, i.e., architecture configuration. 
Several classes and examples can be distinguished for each.
For example, for the temporal constraints, the modeler could specify periods in which the policy is active or could specify \textit{cooldown} periods after an adjustment is made.
For \textit{state-based} constraints, the modeler could specify preconditions and postconditions on the state of the target group, the trigger, or the sensed environments. 

Since multiple policies can be defined for a single target group another attribute for constraints is defining the application scope.
The \textit{scope} determines whether a modeled constraint applies to one policy only or whether the constraint is independent of the policies and applies to the target group instead.
For example, the maximum number of instances could be meaningful in both cases. It can specify the maximum number of instances a policy can create, but it can also specify the maximum number of instances throughout the lifetime of a target group.
Allowing or prohibiting this freedom to the modeler is a design choice.

The third dimension constitutes the \textit{behavior}, which we classify into two subclasses: \textit{prohibiting} and \textit{altering}.
The former aligns with constraints as specified in the Object Constraint Language (OCL) that are side-effect free~\cite{ocl2014} and are preconditional propositions for the policy to make an adjustment. 
For example, a \enquote{cooldown time period} constraint acts as a preconditional proposition in case the latest adjustment sourcing from the policy is made within the cooldown period, then simply no new adjustment is made.
The other type of behavior is the one which \textit{alters} the decision.
Contrary to the previous example, the \enquote{maximum number of instances} constraint allows the possibility to alter the adjustment.
For example, if the intended number of elements is greater than the defined maximum number of elements, then the target group shall scale up to the maximum number of elements instead of forbidding that adjustment entirely.

\subsection{Visual Notation}
To foster understanding and intuition behind our modeling language, we devise a graphical syntax following the requirements from Moody~\cite{Moody2009} for visual notation in software engineering. 
The notation adheres to Moodys requirements~\cite{Moody2009} by fulfilling the following defined requirements: (1)~having the design goal of representing elasticity policies visually to aid communication and comprehension, (2)~having a one-to-one mapping between the metamodel and the graphical notation, (3)~having a high visual distance between elements, (4)~differentiating with text labels element instances, (5)~using notations that resemble the meaning of the concepts, and (6)~considering the span of judgment of the human brain is roughly seven~\cite{Miller_1956} the notation should not contain more than five to nine elements.
The process and the resulting graphical editor are part of~\cite{Summerer2022}. 
In this section, similarly to the initial graphical editor developed by Summerer~\cite{Summerer2022}, we provide a graphical notation for the elements of SPD. 
We use this notation throughout the examples and we use the same notation in the generated Eclipse Modeling Framework tree editor.

For the main modeling entities, we use shapes as the main visual variable to distinguish concepts in the notation. In addition, we use text labels in the near proximity of the shapes to fully represent a modeling concept.
A target group is represented with a dashed circle to denote that it is a set of elements that can change due to the applied policies.
A scaling policy is noted with a portrait rectangle and a label that defines the name of the scaling policy. 
A square identifies a trigger, and an equilateral triangle notes an adjustment type. In composition, it allows quick identification of the policies as ordered elements that the architect can read naturally from left to right, i.e., "fires based on the trigger and makes a particular adjustment".
For constraints, we use the octagon inspired by the traffic stop sign with a cross in the middle. 

The application of a policy to a target group is explicitly marked with an arrow that links the scaling policy instance 
The other relationships in the model are visually noted by placing elements in the proximity of the containing or parent elements. 
For the trigger and the adjustment type, the scaling policy rectangle acts as a separator and anchor for the trigger symbol (on the left) and the adjustment type symbol (on the right). 
The spatial position of the octagon conveys whether a constraint applies to the policy or the target group. 
We provide additional helpful information through text labels to help the end-user understand the type of elements used and the modeled parameters.
Such a notation is useful for different tools that may be built around the elasticity view type based on the SPD model as well as for communication purposes.

\newpage
\begin{table}[ht]
\centering
\begin{tabular}{| m{1.5cm} | >{\centering\arraybackslash}m{1.5cm} | m{4cm} |}
\hline
\textbf{Element} & \textbf{Symbol} & \textbf{Description} \\ 
\hline
Policy & \includegraphics[width=1.5cm]{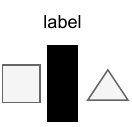} & 
A \texttt{ScalingPolicy} is constrained to contain one trigger and one adjustment. The \texttt{label} maps to the name attribute value of an instance.   \\  \hline
Target & \includegraphics[width=0.8cm]{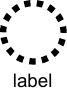} & Represents a \texttt{TargetGroup} instance. The \texttt{label} maps to the name attribute value of an instance.  \\  \hline
Trigger &\includegraphics[width=1.2cm]{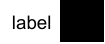} & A \texttt{ScalingTrigger} instance. The \texttt{label} is a text computed based on its constituents constructs, e.g., \enquote{cpuUtil > 80\%}.\\ \hline
Adjustment & \includegraphics[width=1.2cm]{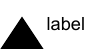} & An \texttt{AdjustmentType} instance.  The \texttt{label} is a text computed based on its type, e.g., \enquote{+1} for a \texttt{StepAdjustment} or \enquote{+10\%} for a \texttt{RelativeAdjustment}. \\  \hline
Constraint & 
\includegraphics[width=0.6cm]{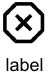}
& A \texttt{Constraint} instance. The \texttt{label} is a text computed based on its type, e.g., \enquote{CD=60s} for a \texttt{CooldownConstraint} \\  \hline \hline
\end{tabular}
\begin{tabular}{| m{2cm} | >{\centering\arraybackslash}m{1cm} | m{4cm} |}
\hline
\textbf{Relation} & \textbf{Symbol} & \textbf{Description} \\ 
\hline
Policy Applying to Target & \includegraphics[width=0.7cm]{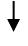} & The arrow links a \texttt{ScalingPolicy} to a single \texttt{TargetGroup}.\\ \hline
Policy containment of a Trigger & N/A & No explicit symbol. The contained trigger is placed on the left side of the policy.\\ \hline
Policy containment of an Adjustment  & N/A & No explicit symbol. The adjustment is placed on the right side of the policy. \\ \hline
Policy containment of Constraints  & N/A & No explicit symbol. Constraints are placed on top of the policy. \\ \hline
Target containment of Constraints  & N/A &  No explicit symbol. Constraints are placed near to the target group. \\ \hline
\end{tabular}
\caption{Visual Notation for the Elasiticty View Type based on SPD.}
\label{tab:notation}
\end{table}

\begin{figure}[ht]
    \centering
    \includegraphics[width=0.3\textwidth]{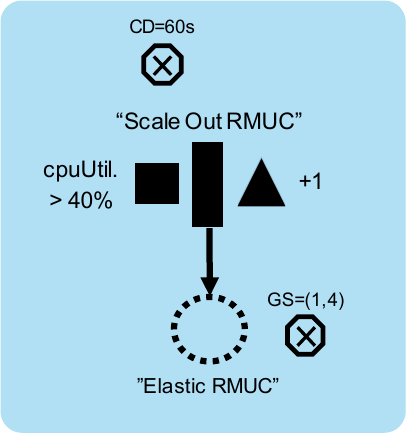}
    \caption{Example SPD Model where a \texttt{ScalingPolicies} is applied to an \texttt{ElasticInfrastructure} target group for the running example.} 
    \label{fig:example-elasticinfrastructure}
\end{figure}

\Cref{tab:notation} depicts the symbols and the description of the notation for all the modeling elements of SPD whereas \Cref{fig:example-elasticinfrastructure} shows an example instance with all the elements of the SPD modeling language used. 
In the example, a \texttt{ScalingPolicy} instance applies to the \texttt{TargetGroup} instance named \enquote{Elastic RMUC}. 

The policy is named \enquote{Scale Out RMUC} and contains a \texttt{SimpleFireOnValue} scaling trigger that bridges the \texttt{CPUUtilization} stimulus and the expectation \texttt{ExpectedPercentage} with a value of \texttt{40\%} and a relational operator \texttt{GreaterThan}. 
The policy uses a \texttt{StepAdjustment} with a step size set to one. 
The policy adds one element to the target group upon firing. 

\newpage
\subsection{Simulation Semantics}
The purpose of this section is to describe conceptually the differentiation between the bottom-up and top-down transformation semantics that distinguish two possible and mutual exclusive ways of modeling through SPD. 
The transformations are implemented in QVTo and, together with the runtime metamodel, are publicly available\footnote{https://github.com/PalladioSimulator/Palladio-Addons-SPD-Metamodel}.
To define the transformation semantics, we initially define a runtime metamodel for SPD that defines the set of valid runtime configurations. 
Then, we define transformations that determine a transition relation between configuration instances.

\begin{figure*}
    \includegraphics[width=1.05\linewidth]{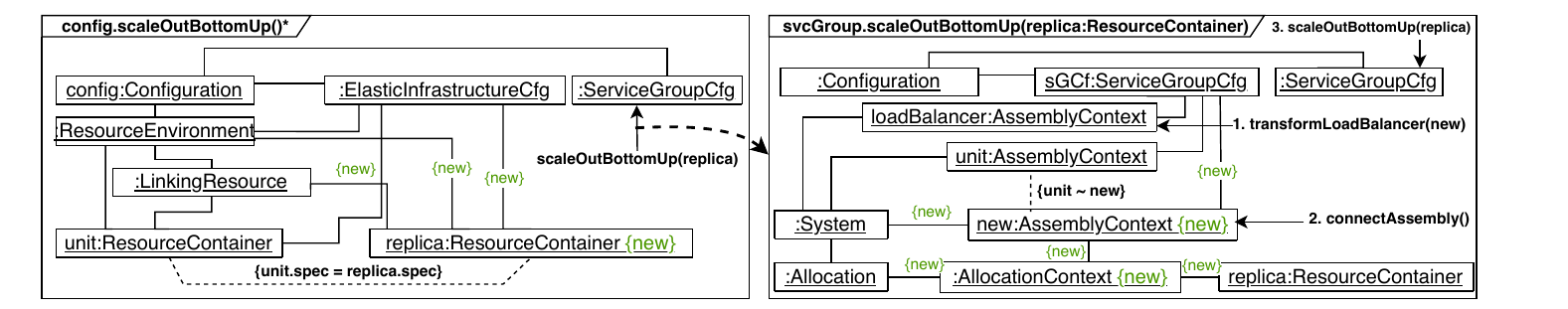}
    \caption{Bottom Up Scaling Out Example Instance to Illustrate the Transformations in the Simulation Semantics. }
    \label{fig:bottom-up}
\end{figure*}


The runtime metamodel contains a root class \texttt{Configurati\-on}, which references all PCM models, including the SPD model. In addition, it points to the currently enacted policy and keeps track of the state of all target groups through a containment to the \texttt{TargetGroupCfg} class. 
Upon enacting a scaling policy from the defined policies in the \texttt{SPD}, at random or during simulation by checking simulation data, the \texttt{Configuration} moves to a new state. 
The abstract class \texttt{TargetGroupCfg} keeps a history of enacted policies during the lifetime of each target group. 
For each target group that is part of the SPD modeling language there is a corresponding concrete configuration class that inherits from the \texttt{TargetGroupCfg}. 
When defining the semantic configuration, the self-adaptive system architect defines the initial configuration of the target groups. 
Independent of whether the modeled and analyzed policies in an SPD model target infrastructure or assemblies, there is exactly one runtime semantic model. 
We name the creation of the runtime semantic model as the Slingshot identification step.
Afterward, the software architect may create SPD models that target services or infrastructure. 
For better comprehension and non-ambiguity in the semantics, we decide to make the types of targets in an SPD model mutually exclusive between service targets and infrastructure targets. 

When the self-adaptive system architect defines policies on infrastructure targets (e.g., \texttt{ElasticInfrastructure}), then the bottom-up transformation semantic defines the transitions of the runtime configuration. 
However, when the policies are defined for the service targets (e.g., \texttt{ServiceGroup}, \texttt{CompetingConsumersGroup}), then the top-down transformation semantics define the transitions. 
In this section we describe the bottom-up and the top-down transformation semantic transformation semantic for a single scale out action to highlight their differences. 
The simulation semantic handles both the scaling out of the target groups as well as the scaling in based on the modeled policies through SPD. 


\paragraph{Bottom-Up Scale-Out.}
The bottom-up scale-out semantics define how the system scales out an \texttt{ElasticInfrastruct\-ure} by increasing its resources in response to a policy defined in the SPD model. When such a policy is triggered, the infrastructure scales by adding more elements, as illustrated in \Cref{fig:example-elasticinfrastructure}. This approach is termed "bottom-up" because the scaling begins at the infrastructure level, and the software components allocated on top of the infrastructure are scaled as a consequence.

The transformation targets the corresponding \texttt{ElasticIn\-frastructureCfg} associated with the triggered policy. Specifically, it adds a finite number of \texttt{ResourceContainers}, based on the type of adjustment and the current number of elements (c.f., \Cref{sec:palladio}). \Cref{fig:bottom-up} exemlifies the transformation, where the size of the target group is increased by one instance. The left side of the figure shows how the new resource container is linked to the existing resource environment, inheriting the same processing specification as the unit resource container. The right side shows the subsequent step, where the number of assembly replicas is increased for each service allocated on the \texttt{ElasticInfrastructureCfg}, using the transformation rule \textit{scaleOutBottomUp(replica)}.

The right side of \Cref{fig:bottom-up} also illustrates the transformation of a \texttt{ServiceGroupCfg}, where a new assembly context is allocated to the newly created resource container. For this to occur, the runtime metamodel must include an elastic configuration, such as a \texttt{ServiceGroupCfg} or \texttt{CompetingConsumersGroupCfg}, which enables the automatic scaling of the assemblies.

Both transformation rules correctly replicate the newly created elements. 
In the left rule, the newly created resource container must match the processing specification of the unit resource container. 
In the right rule, the new assembly context must align with the unit's behavior and be properly connected to the load balancer assembly. 
For the service group to handle the scaled-out load, it relies on the existence of an additional component in the repository and a corresponding assembly context instance. The transformation process is encapsulated in the subsequent steps: \textit{1. transformLoadBalancer(new)} and \textit{2. connectAssembly()}.


\begin{figure*}
    \centering
    \includegraphics[width=\linewidth]{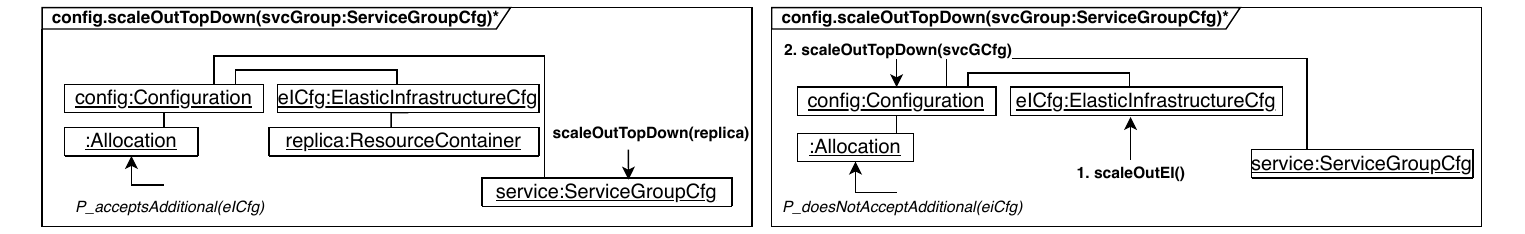}
    \caption{Top Down Scaling Out Example Instance to Illustrate the Transformations in the Simulation Semantics. }
    \label{fig:topdown}
\end{figure*}

\paragraph{Top-Down Scale-Out.}
Assuming the enactment of a scaling policy applied to one of the service targets, either \texttt{ServiceGroup} or \texttt{CompetingConsumersGroup}, a top-down transformation modifies the \texttt{Configuration} instance accordingly.
In a top-down scale-out, the matching target group configuration is transformed by creating a new replica and allocating it to one of the resource containers in the elastic infrastructure.

A key distinction from the bottom-up transformation is that not all service groups are scaled out, but only the target service group, and the infrastructure is transformed as needed.
For the top-down scale-out semantic, the maximum number of allocatable service replicas per resource container must be predefined.
A post-condition is that all resource containers, that are part of the elastic infrastructure, must not exceed the maximum number of allocatable service replicas, and all newly created assemblies are allocated to resource containers within the \texttt{ElasticInfrastructureCfg}.

\Cref{fig:topdown} illustrates the transformation of a single service group target.
The rule on the left demonstrates the case where the current elastic infrastructure configuration can allocate a newly created assembly (reflected by the predicate \textit{P\_acceptsAdditional(eiCfg)}). Additionally, it shows the matching of the resource container named \enquote{replica}, which is the hosting container of the newly created service assembly.
The rule on the right presents the alternative case, where no resource container exists that can host an additional replica. In this case, the elastic infrastructure is scaled out to accommodate the new service replica (\texttt{scaleOutEI()}).
Upon the transformation of the infrastructure with new resource container(s)\footnote{Based on the policy, there may be more than one replica instantiated, requiring multiple resource containers.}, the service target transformation follows the same process as in the bottom-up semantic (cf., \Cref{fig:lts}, right part); the main differences lie in the consequences and continuation of the transformation.
In the bottom-up semantic, changes occur as a consequence of the elastic infrastructure being scaled, whereas in the top-down approach, the transformation drives the adaptation of the infrastructure.
Moreover, in the top-down transformation, the process terminates after transforming the target group to which the policy applies, whereas in the bottom-up transformation, all service groups within the elastic infrastructure are transformed.

The description of the semantics in this work aims to illustrate the distinction between how the configuration of the architecture evolves under two different types of policies.
However, it does not fully cover all the features of the transformation semantics. 
The simulation semantics also addresses scaling in of architectural elements when policies reduce the number of elements in a target group, i.e., scaling in the architecture as a consequence of applied scaling policies. 
It manages the creation of new connectors and the modification of auxiliary components, such as those with load-balancing capabilities and request routing.
Both semantics share the same configuration, i.e., the mapping of architectural elements to elements on which scaling policies can be applied using SPD. 
The two semantics govern transformation behavior in distinct cases: when policies are defined for elastic infrastructures and when they are defined for services.
The distinction of the two semantics allows the modeling of several use cases and concerns with respect to elasticity while maitining clarity and understandability of their effects.

\section{Evaluation}
The main object of investigation revolves around two dimensions that the proposed elasticity architectural view impacts: predictive power and productivity. 
For predictive power we measure the prediction accuracy in varying cases to determine the validity of the abstractions for making accurate performance predictions for elasticity.
In addition, we investigate the simulation efficiency for elasticity in comparison to load testing as well as with changing the size of the simulated architecture. 


For determining prediction accuracy, we require estimating ground-truth values for a set of performance indicators to compute prediction accuracy. 
We base the estimation of ground truth values on the proposed methodology for performance testing of cloud applications by He et al.~\cite{He2019}. 
The main concern in estimating ground truth values is the intrinsic uncertainty of cloud environments, i.e., incomplete knowledge of the execution environment and changes in between experiments. Hence, the objective is to establish a minimum number of replications to reliably estimate ground truth performance indicators. 
Depending on the cloud environment, this number varies. We rely on the bwCloud\footnote{https://www.bw-cloud.org} to conduct our experiment and establish that for the designed experiment, executing two replications suffices.

We select two cases to study the treatment in varying cases. The first case, the Remote Measuring Use Case (RMUC), defines a small case with characteristics of micro\-service-based applications. The size of the case allows us to determine the prediction accuracy with lower effort. The second case defines a larger business-information microservice-based application (Microservice-BIS) that stems from previous research where we highlight the problem of modeling and prediction elasticity. We could not establish prediction accuracy for the second case because we lacked control and access to an implementation. However, the case allows us to observe the impact of our intervention on larger models. 

In both cases, we vary the style of elasticity policies and their configuration to determine the validity range of the designed architectural view and its corresponding semantics. Through this variation, we aim to show the capabilities and determine the cases in which refinements are required or unsuitable for the approach. 

For productivity, the running claim is that the proposed elasticity architectural view type increases productivity in comparison to existing works that achieve the same semantics by workarounds or through general-purpose modeling languages and enables better design capabilities. 
To gather data on how the view is perceived by self-adaptive software architects, we design and conduct a user study in which we let participants solve two tasks using the proposed architectural view. 
The tasks are of two different types: in the first, participants must model from scratch a scenario in which elasticity policies are defined for a specific cloud provider, and in the second, participants must refine an existing model in the elasticity architectural view. 
The refinement task requires the creation of an additional scaling policy in the architectural view that points to the right service as a remedy for the bottleneck shifting scenario.
Bottleneck shifting is a typical scenario when performance engineering microservice-based architecture concerns elasticity. 
In such a scenario, a service that initially is the bottleneck is autoscaled. However, this does not improve the overall performance, while an upstream service becomes the new bottleneck. Participants are tasked with understanding the scenario, identifying the existing policies in the architectural view, and adding a new one for the right service as a remedy. Both tasks involve domain knowledge and tool knowledge; hence, we conducted a pre-training session to improve their understanding.
During the second session, where participants solve the tasks, we measure time as an objective measure for effort needed to complete them. 
At the very end, we collect all the models constructed during the session to compute the correctness score. Additionally, participants must rate their performance during the session and the tool's usability.  
The collected data allows to internally validate the approach and present a Type~2 validation where end-users other from the creators use the architectural view. 
Further, it allows us to contrast the results with existing empirical evidence to approximately determine the associated effort of the new architectural view.

\subsection{Prediction Accuracy}

\begin{table*}[ht]
\scriptsize
\centering
\begin{tabular}{|l|l|l|l|}
\hline
\textbf{Policy} & \textbf{Style} & \textbf{Semantics} & \textbf{Configuration} \\
\hline
\texttt{nodebased-40} & Centralized, Infrastructure, Plastic & Bottom-Up & Trigger: 1 Min Avg. Node Load > 40\%, Adjustment: +1 Node \\
\texttt{nodebased-60} & Centralized, Infrastructure, Plastic & Bottom-Up & Trigger: 1 Min Avg. Node Load > 60\%, Adjustment: +1 Node \\
\texttt{nodebased-40-E} & Centralized, Infrastructure, Eager, Plastic & Bottom-Up & Trigger: 1 Min Avg. Node Load > 40\%, Adjustment: +2 Nodes \\
\texttt{nodebased-60-E} & Centralized, Infrastructure, Eager, Plastic & Bottom-Up & Trigger: 1 Min Avg. Node Load > 60\%, Adjustment: +2 Nodes \\
\texttt{d-hpa-def} & Decentralized, Services, Elastic & Top-Down & Target tracking CPU util.: 80\%\\
\texttt{d-hpa-def-60} & Decentralized, Services, Elastic & Top-Down & Target tracking CPU util.: 60\% \\
\texttt{d-metrics-ql5-rt0.5} & Decentralized, Services, Elastic & Top-Down & DProc: Target tracking QL 5, DComm: Target tracking RT 500ms \\
\texttt{d-metrics-ql5-rt1} & Decentralized, Services, Elastic & Top-Down & DProc: Target tracking QL 5, DComm: Target tracking RT. 1sec \\
\texttt{d-metrics-ql5-rt0.5-cd60} & Decentralized, Services, Elastic & Top-Down & Same as d-metrics-ql5-rt0.5 but with custom cooldown 60 secs\\
\texttt{d-metrics-ql5-rt1-cd60} & Decentralized, Services, Elastic & Top-Down &  Same as d-metrics-ql5-rt1 but with custom cooldown 60 secs \\
\texttt{max} & Static & - &  \\
\texttt{none} & Static & - &  - \\
\hline
\end{tabular}
\caption{The style and configuration of elasticity policies for the RMUC case.}
\label{tab:merged_policy}
\vspace{-10pt}
\end{table*}

A set of 60 data points constitutes the space of explored policies for determining the prediction accuracy behind the simulation of elasticity using the proposed viewtype. 
The set contains 12 different policies for elasticity and their behavior on five different workload levels with an increasing number of users sending requests.  
The 12 configurations differ in how elasticity of the application is achieved both in terms of style and configuration.
\Cref{tab:merged_policy} shows the style and configuration of the modeled policies.
We conduct load tests for each configuration in an experimental setup consisting of a six-node Kubernetes cluster (v1.26.5) allocated on provisioned virtual machines from bwCloud.  
Each virtual machine has an \texttt{m1.large} flavor with 4VCPUs and 8GB memory.  
Two nodes are excluded from deploying the components of the RMUC. 
One is reserved for the Kubernetes master, and one is reserved for the load generator. 
The prototype utilizes the remaining four nodes. 
We execute 5 simulation replications for each configuration to estimate the interval for performance indicators of interest such as the average response time for devices sending data, throughput, or the 95th percentile of response time (c.f., ~\Cref{sec:scenario}) for a five minute fixed-length interval. 
The objective of the experiment is to determine the accuracy of the simulation model to evaluate the elasticity of the application in a fixed duration interval.
To focus on the adaptivity of the application and the model, instead of modeling the application's performance characteristics, we rely on ProtoCom~\cite{Becker2008} to generate parameterized resource demands similar to the modeled demands in the Palladio model. This way, the Palladio model and the application exhibit similar performance characteristics in a low-load and static architecture configuration.   

\begin{figure}[ht]
    \centering
    
    \includegraphics[width=0.5\textwidth]{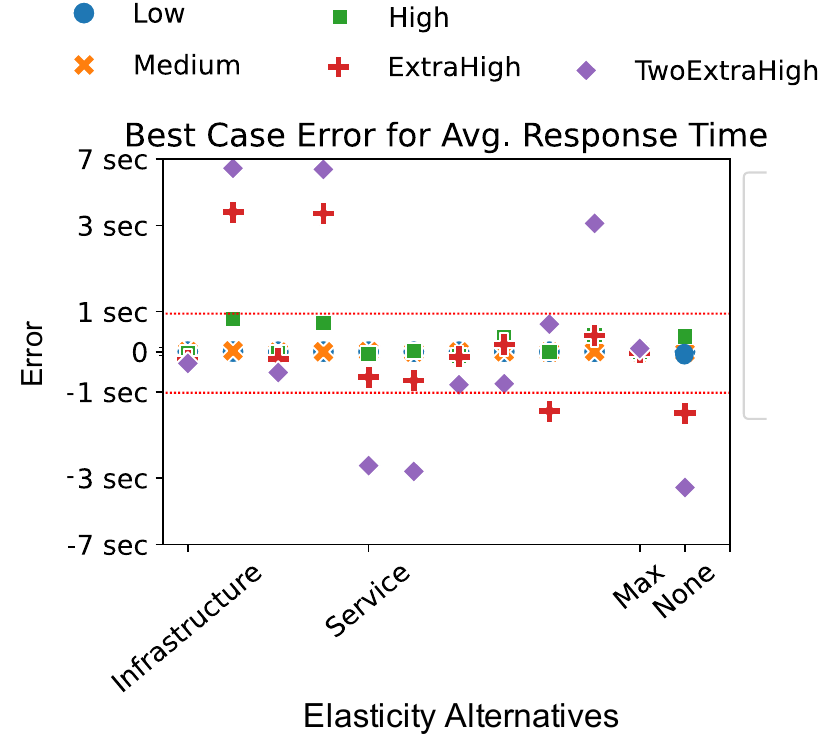}
    \caption{Overall accuracy for Average Response Time under different policies for the RMUC under different workload levels.}
    \label{fig:results-all}
    \vspace{-10pt}
\end{figure}

\Cref{fig:results-all} shows best case error for the average response time of devices sending data across workload levels for different elasticity policies.
The x-axis shows the policies where only the first policy within a group of policies is labelled (i.e., Infrastructure (\texttt{nodebased-40}), Service (\texttt{d-hpa-def}), Max (\texttt{max}) and None (\texttt{none})). 
The y-axis shows the error magnitude in seconds and each datapoint is represented in a five different shapes to indicate the different workload level.
One datapoint shows the smallest distance of the estimated average response time by means of load testing to the estimated interval by means of simulations for one out of the twelve policies shown in \Cref{tab:merged_policy}. 
If the best case error is bounded to a predefined value of one second we classify the prediction as valid and we include the case to the set of cases for which we calculate the mean absolute percentage error. 
If the case is considered invalid refinements to the model have to be made based on hypothesized causes that lead to the large deviation.
We base the analysis on the method suggested by Banks for validaation of simulation models~\cite{Banks2004DiscreteEvent}.

Overall, predictions get worse for scenarios with a higher number of concurrent devices sending data, i.e., ExtraHigh and TwoExtraHigh indicating the inability to capture aspects of the elasticity for those cases. 

The first four policies depicted in the x-axis define elasticity policies in which the system exhibits a degree of plasticity where the application is scaled out, but no policy is defined to scale it back into its initial state.
Due to space restrictions, we label only the first of the category with the name \texttt{Infrastructure}.  
Configurations in the category differ by the CPU utilization threshold used to spin up a new application instance that consists of the three application components of the RMUC case and the number of application instances to spin. 
The first and third policies use a 40\% CPU utilization threshold, whereas the second and the fourth use a 60\% threshold. 
Further, within the same tuples, one of the variants is eager, where the adjustment is set to 2, meaning that two more instances are created upon scaling.
The 60\% threshold elasticity policies did not trigger adaptations in the experimental setup, contrary to the model in which adaptations were present. 

In case of two configurations where the threshold is set to 60\% for scaling the infrastructure the error exceeds by far the predefined threshold of one second. In the two configurations, the simulator falsely assumes that there are scaling decisions, however, they do not occur in the load tests. 
Hence, the simulator overestimates performance by predicting a lower value for the average response time that differs by around three seconds in the ExtraHigh workload scenario and around seven seconds in the TwoExtraHigh workload scenario. 

\begin{figure*}[h]
    \centering
    \includegraphics[width=0.3\textwidth]{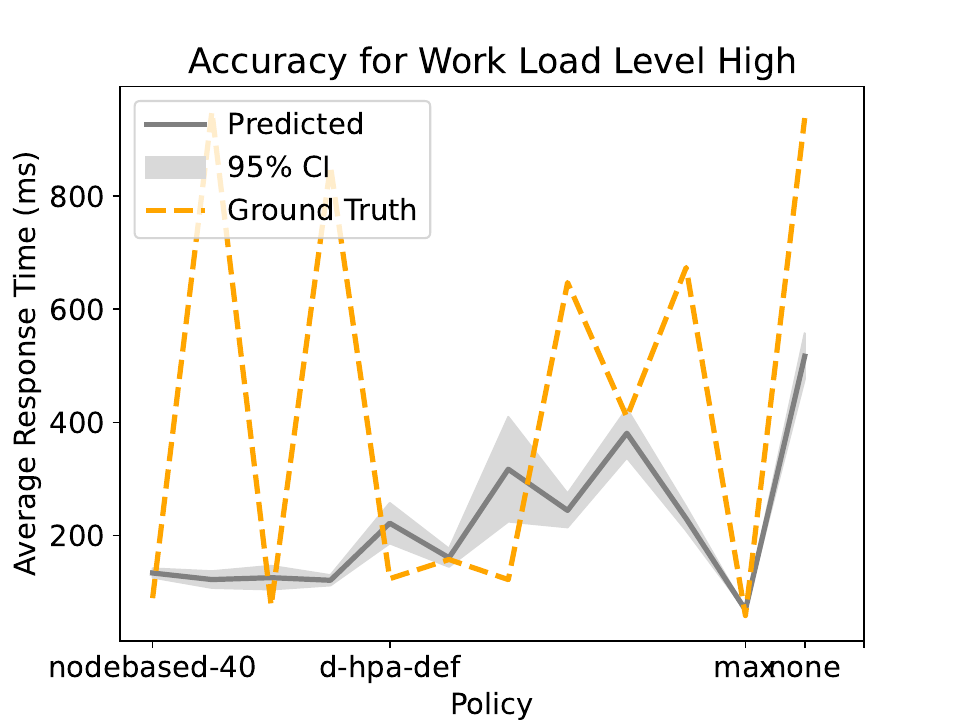}
    \includegraphics[width=0.3\textwidth]{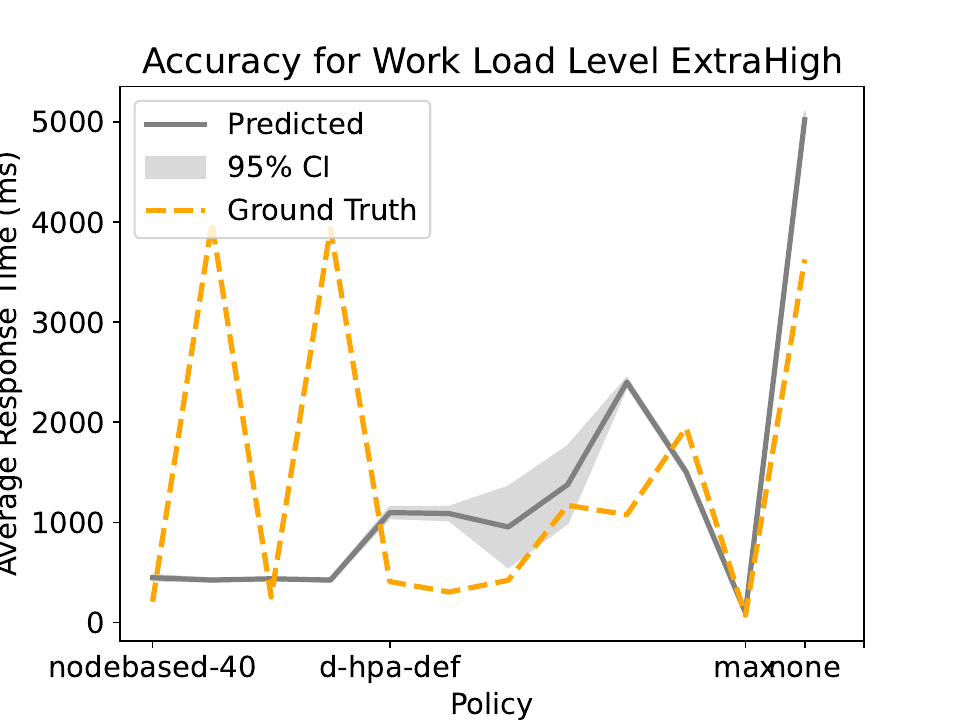}
    \caption{Accuracy for Workload Levels High and ExtraHigh for the RMUC.}
    \label{fig:per-workload}
\end{figure*}

In the service based cateogory, the first two configurations map to the behavior of the Kubernetes HPA \cite{kubernetes_2022} on which target tracking policies are employed. 
In the ExtraHigh, the difference is more severe. 
The simulated policy contains fixed adjustments where one replica is added when the measured CPU utilization is exceeding the defined target (and one is removed when it is belowed the target). However, in HPA, the adjustment magnitude are trigger dependent. In case the measured CPU utilization is exceeding by a factor of two the set target then two additional replicas are created. Hence, in the first two service-based configurations, the simulation is not able to capture the right adjustment magnitude and predicts higher response times by around three seconds.  

In the 4 follow-up policies application components of the RMUC have their own scaling policies defined, and they scale independently out and in. 
The error for the first two configuration where the cooldown period is set to 180 seconds is smaller in comparison to the two configuration in which the cooldown period is set to 60 seconds with a higher dynamism in which more scaling decisions are permissable.





The last two configurations at the end, namely Max and None define static policies in which the architecture is simulated at a static configuration with maximum number of resources and with a minimum number of resources. 
As the figure shows, in the Max configuration, the prediction error is relatively smaller in comparison to the None case and to the other cases with scaling present. 


\Cref{fig:per-workload} depicts the accuracy of predictions for two workload levels, High and ExtraHigh. 
In both workload levels, predictions follow the shape of the collected ground truth data. An exception is the case where a node-based elasticity policy is employed with a 60\% threshold trigger, as discussed previously. 
The gray area depicts the 95\% confidence interval of 5 simulation replications. 
Policies with more architectural dynamism, such as the decentralized ones, have a larger CI width than the static and plastic elasticity policies. Such configurations require more simulation runs to make the interval estimation narrower. 

\paragraph{Prediction Error.}
The RAE defines the relative difference observed between a ground-truth value and estimated value for an indicator such as the average response time. 
A value of 30\% means that the difference between the prediction and the ground truth estimate is 30\% of the actual value.
For example, if the prediction is 1 second and the ground truth measurement is 2 seconds, the resulting RAE is 50\%. 
The RAE metric is relative to the prediction; a higher RAE does not necessarily mean a higher difference in seconds across cases. 

The resulting best-case and worst-case mean absolute percentage error (MAPE) of predictions is 27\% and 46\% if we remove five outliers that significantly contribute to the large mean (>150\% RAE). So, the best-case MAPE of 27\% for the filtered data points is acceptable considering the established criteria and the worst-case MAPE of 46\% is close. 
However, if all the data points are considered, the best-case and worst-case MAPE worsens to 52\%, respectively, 73\%. 
In the worst case, most predictions (86\%) have residuals of less than one second compared to the ground truth estimates.
The predictions differ by more than one second from the ground-truth average response time in seven data points. 
The policies in which such discrepancies occur are the None case~(2), the policies that mimic the HPA default behavior~(2), and the policies based on application metrics with the cooldown set to one minute~(3). 
In the best case, predictions do not improve significantly regarding the absolute error and the set threshold of one second.

\paragraph{Correlation of Speedups for Other Performance Indicators.}
We compute the speedups in the performance metrics compared to the case where no policy is employed. We use the latter as a baseline. 
The predicted speedups show a strong positive correlation with ground truth speedups for average response time ($r=0.89, p=2.72\times10^{-16}$) and throughput ($r=0.94,p=5.3\times10^{-21}$), but only a moderate positive correlation when it comes to the tail latencies reflected by the 95th percentile ($r=0.61, p=7.10\times10^{-6}$).

This brings evidence that the predicted performance of various elasticity alternatives using a Palladio model and the SPD model is likely to correlate to the ground truth when considering the three metrics: the average response time, the throughput, and the 95th percentile of response times. 

\begin{table*}[ht]
\centering
\small
\begin{tabular}{|l|l|l|l|l|l|l|l|l|l|}
\hline
& \multicolumn{4}{c|}{T1 (Creating SPD from scratch)} & \multicolumn{4}{c|}{T2 (Alleviating Bottleneck Shifting)} & \multirow{2}{*}{SUS} \\ \cline{1-9}
Participant & Time (min) & Size & Size/Time & Correct. & Time (min) & Size & Size/Time & Correct. &  \\ \hline
P1 & 31.00 & 45.52 & 1.47 & 0,96 & 21.00 & 22.36 & 1.06 & 0,98 & 87,5 \\  \hline
P2 & 24.00 & 37.32 & 1.56 & 0,6 & 29.00 & 32.20 & 1.11 & 1 & 77,5 \\ \hline
P3 & 32.00 & 25.88 & 0.81 & 0,64 & 28.00 & - & - & 0 & 62,5 \\ \hline
P4 &  13.00 & 45.52 & 3.50 & 0,904 &  10.00 & 19.18 & 1.92 & 0,98 & 90 \\ \hline
P5 & 28.00 & 37.04 & 1.32 & 0,8 & 25.00 & 19.18 & 0.77  & 0,6 & 45 \\ \hline
P6 & 18.00 & 45.52 & 2.53 & 0,704 & 20.00 & 19.18 & 0.96 & 0,98 &  60 \\ \hline
\end{tabular}
\caption{Time spent modeling with SPD and the computed model size for six participants for T1 and T2, their correctness score, and the overall rated system usability score (SUS)~\cite{brooke1996sus}.}
\label{tab:productivity}
\end{table*}

\subsection{Simulation Efficiency}
The first case allows us to capture the prediction accuracy. However, the low number of microservices does not represent modern microservice-based applications with respect to size~\cite{newman2021building}. Hence, we additionally validate the approach with a second case of a larger size~\cite{Klinaku2021}. 
The simulation experiments were conducted on an off-the-shelf PC for both cases, demonstrating reproducible results using commercially available and inexpensive hardware.
The benefit of simulating the first case compared to load testing the elasticity policies in the cluster is two orders of magnitude less overall computation time, i.e., we effectively spend 120 hours of execution time determining the system's behavior in five minutes for the varying closed workloads. 
However, the simulation time increases by an order of magnitude for the larger case. Further, the simulation time increases by almost a factor of three in the first case and four in the second case between the elastic configurations compared to static.

\subsection{Modeling Efficiency} 
In two sessions of 90 minutes six participants of similar background participated in the study. Five participants were postgraduate students, and one was a doctoral-level student. Three participants finished their bachelor's degree in Europe, and the other half in Asia, but all are undergoing studies in Germany. 
Participants achieve an overall average correctness of 76\% with a standard deviation of 28\%. Further, they spend, on average, 23.25 minutes solving each task.  
On average, participants perform better in the first than in the second task.  
Concerning task requirements, in the first task, participants receive fewer points in the target group and type of adjustment. 
Two participants could not model the right adjustment type in the first task, according to the scoring scheme, the participants modeled adjustments that did not increase the capacity. One participant modeled the wrong target type: a \verb|ServiceGroup| instead of an \verb|ElasticInfrastructure|. 

Further, participants provided subjective perceptions of the difficulty, time needed to solve the task, correctness, amount of work, and assistance with the modeling approach. For each task, participants rated these dimensions on a seven-point scale. 
Related to the difficulty, the tasks have been regarded as easy. One exception exists, where one participant scored both tasks as very difficult.  Regarding time performance, opinions are more diverse and distributed over neutrality, and neither very slow nor very fast scores exist. 
Participants scored higher on correctness for the first task than for the second. One outlier exists; the participant scored correctness as completely wrong. This matches the exception with the rated difficulty and is consistent, i.e., the same participant scored both tasks as very difficult and completely wrong.
Regarding the amount of work, opinions are distributed over neutrality, and no participant perceived the amount of work as too little or too much. 
Last but not least, related to the modeling approach, the majority perceived the approach as helpful for solving the tasks. There are two votes for the extreme option of \enquote{it was helping} for the second task. These two votes are the only ones that mark the highest score in all the questions.

Participants liked the simplified creation of scaling policies and the visual overview that is quicker and more comprehensible than traditional textual descriptions. 
One participant liked the vocabulary alignment to concepts from cloud providers.  One participant appreciated the level of abstraction, and another expressed appreciation for the approach's capability to view a system from different viewpoints.
One statement is that the training session and the demo task were good.  

Participants emphasized several aspects that they did not like. 
The majority of participants state tool-associated issues and their associated complexity. Issues include editor issues, i.e., UI not working, bugs, and the effort to set up Eclipse IDE and navigate the graphic interface, such as dragging model elements to their containers or scrolling within the palette.
Several statements point to the abstraction level of the model: one struggles with how it maps to specific constructs from cloud providers, e.g., evaluation periods of triggers; one is concerned with how the decisions map to configurations later on, and one states the need for additional documentation for constraints. 
Moreover, one user expressed confusion regarding the expected solution and assumptions for the second task, which aligns with our expectation that task 2 involves a more complicated domain problem of finding a remedy for bottleneck shifting to an upstream service.

To estimate the modeling throughput of our view, we compute the fraction of size and time and compare that to previous empirical evidence. Specifically, Martens et al.~\cite{Martens2008} compare the effort of PCM against the SPE method through a controlled experiment. 
We select Halstead's metrics for size computation to compute the size. Halstead proposes four primary metrics: the number of operators and operands in the program (N1, N2) and unique operators and operands (n1,n2).
In our case for operators, we count the number of modeling elements used. For operands, we count the number of attributes. The rationale is that in the modeling activity, two elementary operations exist: constructing model elements and defining their attributes. 
\Cref{tab:productivity} shows the resulting Size/Time ratio for each participant.
We derive approximated and normalized modeling throughputs for the other views of Palladio from~\cite{Martens2008}.
For the given dataset, the SPD viewpoint has significantly lower modeling throughput than approximated throughputs for modeling control flows, resource environment, and usage profile in Palladio. 
However, it is still better than the most tedious task of modeling resource demands. 
When modeling resource demands, a lot of time is spent on a few modeling elements.

\section{Discussion and Limitations}
In this work, we contribute a viewtype that frames elasticity concerns, evaluate the accuracy of performance predictions for elasticity models in a single case experiment, assess the simulation's efficiency for elasticity management, and investigate how end-users model systems using the proposed approach. 
In this section, we discuss the key results and limitations of our contribution. 

The resulting metamodel of the SPD language has been strongly influenced by constructs commonly found in existing cloud providers and related literature.
As a result, SPD supports the modeling of reactive elasticity policies, though it offers limited support for other resource management paradigms, such as reinforcement learning, control theory, or queueing theory-based approaches. However, the core concepts and relationships within the modeling language were intentionally designed to facilitate future extensions.
Ongoing work is focused on extending the metamodel to support the definition of policies with learning capabilities. Furthermore, there is additional evidence of the usefulness and broader adoption of SPD beyond its initially intended use cases.
For instance, Stie{\ss} et al.~\cite{Stiess2022} utilize SPD in conjunction with Palladio to address the challenge of designing coordinated and explainable self-adaptive systems.

We evaluate the ability of the model to make accurate performance predictions by modeling elasticity policies using the proposed view type for a single case that utilizes modern technologies like  Kubernetes and autoscaling features.
The proposed view type abstracts away several aspects which may influence the accuracy. 
The view type does not offer the ability to model the structure and distribution of the adaptation processes, instead, we propose the modeling of policies for transforming architectural elements based on triggers, constraints, and adjustment types.
In addition, we abstract away the resource demands for scaling decisions, meaning that scaling actions are instantenous and happen whenever the modeled trigger fires and the modeled constraints are met. 
Such decisions which make the view more intuitive from end users may severely impact the accuracy. 
Therefore, we investigate how far the predictions are from the more reliable approach of load testing elasticity policies. 
From a design decision-making perspective, the main interest is keeping the error within 30\% for performance predictions.  
A deviation by 30\% has been considered acceptable in literature for making design time decisions~\cite{becker2013simulizar, Amiri2020}. Results yield a mean absolute percentage error of 27\% in the best case  and 46\% in the worst case for valid simulation cases. 
The resulting accuracy both in the best case and the inaccuracy in the worst case can not be attributed entirely to the assumptions made for the view type. 
More comparative evidence is needed to directly test the influence of the assumptions.
We conclude that the view type increases the predictive power of Palladio by allowing the simulation of dynamic software architectures based on intuitive policies that control the number of architectural elements during simulation time.
The accuracy results show that the taken modeling approach provides precision but insufficient accuracy. 
The strong correlation of speedups for average response time, throughput and moderate for the 95th percentile suggests the precision in mapping the configurations to their actual ground truth ranges, however, the accuracy within each configuration is still low for the worst case. 
Since, achieving a desired accuracy level is often requires iterations, the results prove the correct implementation of the transformations and envisioned approach behind SPD. 

The experiments in the two cases highlight the utility of relying on discrete event simulations for elasticity in contrast to performing load tests. 
The simulation increases for larger models suggesting the need for optimizations for larger systems and for longer simulations. 
In our experiments, we rely on terminating simulations~\cite{Banks2004DiscreteEvent} and do not investigate the efficiency for steady state simulations. 

The proposed architectural view type relies foundationally on policy-driven management for distributed systems~\cite{DBLP:journals/jnsm/Sloman94}. 
The proposed language defines a domain-specific language for self-adaptivity engineering of elasticity. This is viewed as desired by Weyns~\cite{weyns2020introduction} with the argumentation that domain-specific languages may aid communication, productivity, and reuse in comparison to using generic means of defining self-adaptivity concerns. 
This aspect aligns with Luckey taxonomy~\cite{luckey2014a} of whether the specification of adaptation mechanisms is \textit{generic} or \textit{domain-specific}. 
While generic approaches exist like EUREMA~\cite{vogel2014model} for specifying adaptivity in general in software engineering, Weyns emphasizes the need for exploring domain-specific approaches \cite{weyns2020introduction}.

Through the user study we collect evidence on the use of the proposed constructs by a representative sample of novice software architects. 
The sample is low to make objective conclusions about the impact of the proposed approach on constructs such as productivity. 
We triangulate the obtained results with previous work that collect empirical evidence to compare the modeling throughput of the new view type with existing view types to model control flow, resource environment, usage profile and resource demands.
Although this comparison helps contrast the study's results, more evidence is required to draw conclusions about the productivity against existing modeling view types in Palladio or view types from other architectural languages. 
A lower modeling throughput for the proposed view type is observed in comparison to control flows and resource environment.
The lower throughput is inconclusive since it may be attributed to the smaller model size or may be attributed to the complexity of the domain itself for the specification of the dynamism of the software architecture. 
Overall, the collected evidence for productivity indicates a positive impact on productivity.
The threat of concluding the positive impact from a low sample size exists.  However, when also considering the sample of users in the user study from Summerer~\cite{Summerer2022}, where 9 software engineering researchers and practitioners with participation from the industry rate positively the graphical editor for SPD and see the diverse use of SPD, it increases the confidence that the treatment achieves the goal and lessens the threat.

\section{Related Works}

After reviewing related work, observing their model type, level of abstraction related to architecture, and supported variability related to the elasticity policies, we make the following observations and conclusions: (1) the majority of approaches are generic and come with a high effort to vary elasticity designs thereby hampering productivity and predictive power, (2) approaches closer to architectural level elasticity modeling merely realize elasticity policy instances and offer the ability to vary configurations but not elasticity policies of different style; the configurations encapsulated in a few parameters are often ambiguous and imprecise when defining the adaptations to the architectural models, (3) the existing modeling languages for elasticity policies mainly target the enactment at runtime and do not facilitate design time analysis, although there exists one instance that aims at relying on such languages also for design time analysis, (4) elasticity is not targeted specifically by most approaches, hence, the provided feedback to the architect is inadequate and hinders the design and decision making for elasticity policies.

The majority of related work approaches use generic modeling and programming languages to perform design-time analysis of elasticity~\cite{DBLP:journals/simpra/VondraS17, DBLP:journals/simpra/AslanpourTTG21,Amziani2013}. 
We distinguish two classes of approaches for generic modeling of elasticity mechanisms for design-time analysis.
The first class of related work approaches, classified as generic, model particular instances in languages and formalisms suitable for analysis, for example, Timed Arc Petri Nets~\cite{Amziani2013} or Queueing Networks~\cite{DBLP:journals/simpra/VondraS17}.
Approaches of the second class rely on general-purpose languages for specifying adaptations of interests, such as Java, QVTo, Henshin, and others. 
A commonality of both approaches is the high effort to evaluate design alternatives for elasticity.
To check alternative elasticity policies of different styles, the initial models in general-purpose languages and formalisms have limited reuse. 

For representations, i.e., means and constructs (i.e., linguistic abstractions), to model the adaptation process(es)---the managing aspect of the cloud-native application (c.f., \Cref{fig:conceptual-model-cna}), we identify three broad classes: (i)~works that rely on existing general-purpose (modeling) languages to model elasticity~\cite{calheiros2011cloudsim, DBLP:journals/simpra/AslanpourTTG21} (ii)~works that rely on analysis and formal models to model elasticity~\cite{DBLP:journals/fgcs/EvangelidisPB18}, (iii)~works that contribute languages and metamodels for elasticity modeling~\cite{DBLP:journals/fgcs/JradBT19, Brabra2020, AlDhuraibi2021}. 

For pragmatics, the main aspect is the goal and intention behind modeling, i.e., why authors model elasticity. 
The approaches in related work exhibit a wide range of objectives that span from predicting Quality of Service (e.g., performance)~\cite{calheiros2011cloudsim, Stier2018}, generating code~\cite{Caglar2013}, runtime enactment~\cite{Morn2011, Brabra2020}, model checking~\cite{DBLP:journals/fgcs/EvangelidisPB18}, or knowledge reuse~\cite{Weerasiri2016}. 

Related works differ also in terms of applications they target and context. 
A majority of approaches aim at domain independence to cover a large class of applications, such as component-based software systems~\cite{VonMassow2011, Araujo‐de‐oliveira2021, Scheerer2020} or microservice architectures~\cite{Frank2022}. 
Others target specific applications and domains, such as data-driven applications~\cite{Alipour2016} or web applications~\cite{Stier2018a}. 
Most found related work focuses on design-time support. The other set of approaches focuses on offering support for the runtime management of resources and specification of policies at deployment time.

Copil et al.~\cite{Copil2013} define a grammar for the Simple Yet Beautiful (SYBL) language, to specify elasticity strategies for cloud applications. 
Jrad et al. propose the Elasticity Strategy Description Language (STRAT)~\cite{Jrad2016a, DBLP:journals/fgcs/JradBT19}.
STRAT is defined through a Backus-Naur form (BNF) grammar. 
An elasticity strategy is constructed through the initialization of sets and set of actions.
The possible actions, given as language constructs in the grammar, are the duplication, the consolidation and the routing of named resources.
In both instances, STRAT and SYBL, the authors aim to interpret the programs at runtime and issue directives to the cloud provider to provision and release resources.

Abbasipour et al.~\cite{Abbasipour2018} propose a specific metamodel for defining elasticity rules.  
The work differs in pragmatics by aiming at generating elasticity rules instead of design-time elasticity analysis.
They rely on the Object Constraint Language (OCL)~\cite{ocl2014} to define an elasticity rule's actions, conditions, and prerequisites. 
Hence, the software architect must understand the relationships constrained by the metamodel and be able to express the logic of scaling in OCL.

\paragraph{Elasticity Simulators.}
CloudSim~\cite{calheiros2011cloudsim} relies on Java to model cloud resources and the associated features.
To simulate specific autoscaling policies, the codebase of CloudSim has been altered and extended. 
For example, Vondra et al.~\cite{Vondra2017} extend the capability of CloudSim for adding a finite number of virtual machines during the simulation. 
The effort associated with programmatically configuring CloudSim has been tackled in~\cite{Caglar2013}. \citeauthor{Caglar2013}~\cite{Caglar2013} propose a DSML and generate the CloudSim program.  Through the language, they propose a higher-level abstraction called \enquote{CloudDeployment} which encapsulates actions, for instance creation, running an application, waiting for startup etc. 

Jawaddi et al.~\cite{Jawaddi2023} propose a simulator for evaluating and designing autoscaling policies based on reinforcement learning (c.f., one of the main classes of autoscaling). 
The authors show experimental results demonstrating the approach's capability to learn and optimize decision-making for policies based on reinforcement learning.
Related to the modeling constructs, the approach relies on parameters that must be programmatically defined for all the components. 
Including such policies into the proposed SPD metamodel to enable the simulation of reinforcement learning policies through the proposed view type is work in progress and not yet supported.

Aslanpour et al.~\cite{DBLP:journals/simpra/AslanpourTTG21} propose AutoScaleSim, a simulation toolkit for auto-scaling of cloud applications.
The specification of an autoscaling policy in the case of AutoScaleSim~\cite{DBLP:journals/simpra/AslanpourTTG21} is realized through a set of Java classes that follow the MAPE-K pattern~\cite{Computing2006}. 

Vondra et al.~\cite{DBLP:journals/simpra/VondraS17} take a different approach. 
They rely on the R language as a host language to model autoscaling policies. 
As such, the different policies are encoded as R functions. 
They model various policies that differ based on the trigger: response-time-, queue length- or utilization-based trigger. 
Vondra et al. \cite{DBLP:journals/simpra/VondraS17} uses a queueing network model to evaluate the performance of the application. 
Since solving the queuing network model yields steady-state performance measures, they discretize the input time in small enough steps (i.e., 15 minutes) and assume that in each step, the system is statistically steady.
We rely on discrete event simulations for analysis and focus on providing an adequate modeling view type for elasticity of the software architecture. 
It is subject of investigation to determine which approach yields a higher efficiency both for the analysis time as well as productivity.





\paragraph{Architecture-based Elasticity Analysis.}Related work approaches that support modeling and analysis of elasticity behavior at the architecture level merely allow defining configurations of particular realizations~\cite{DBLP:journals/spe/LehrigHB18,Frank2022}. 
Such configurations do not define policies. Instead, they allow the definition of configuration parameters for a realized policy in the platform. 
Hence, they are imprecise, ambiguous, and hard to comprehend. 

In the case of the combination of SYBL and Palladio~\cite{Araujo‐de‐oliveira2021}, the SYBL definitions are meant for controlling elasticity at runtime; hence, they are verbose. 
In addition to that, from the definition itself, several ambiguities arise. 
The adaptation actions must either be modeled explicitly in a transformation rule or an existing transformation has to be reused and referenced in SYBL, as is the case with \texttt{ScaleUp} and \texttt{ScaleDown} references.  
Architectural adaptations are again hidden in transformation scripts, and the end-user in the architect's role is unaware of the policy's effects on the architecture.



In the Palladio ecosystem, we distinguish the following variants of expressing adaptivity concerns: intrusive, invasive, none of the two or certain grades of invasion and intrusion. 
Annotating existing views with templates that have adaptivity concerns can be classified as intrusive. This is the case with the Architectural Template method that has the main goal to capture and reuse architectural patterns \cite{DBLP:journals/spe/LehrigHB18}. 
Using existing modeling constructs to model adaptivity concerns can be classified as invasive. 
An invasive approach has been taken by Stier et al.~\cite{Stier2017} where adaptivity is modeled in conjunction with the domain. The scaling of resources is not an instantaneous actions and their execution consumes computational resources, hence, they consider these performance influencing factor in the model and analysis. 
The proposed approach aims to be both non-intrusive and non-invasive with a separate view that allows solely the expression of adaptivity concerns of scaling. 
This aligns with Luckey's~\cite{luckey2014a} taxonomy of whether the specification of adaptation mechanisms is \textit{implicit} or \textit{explicit} and in case it is explicit, whether it is \textit{imperative} or \textit{declarative}.
In the proposed architecture view, the modeling is made explicit and declarative. 

\textit{Software Architecture View Types and Viewpoints.} 
The paper contributes a dedicated view type that can be classified as one view type in the viewpoint of self-adaptivity or reflection for software systems. 
In classical view models such as the 4+1 view model from Kruchten~\cite{469759} there are views proposed to capture the allocation of logical software components to hardware resources. 
However, for cloud applications the structure changes over time. 
The Unified Modelling Language (UML)~\cite{Rumbaugh2004} is a prominent example of the multipurpose use of models in software engineering.
For describing dynamic aspects of the system, UML provides three views: the state machine view, the activity view, and the interaction view. 
In the state machine view, for example, end users can construct state machine diagrams using concepts such as completion transition, do activity, effect, event, region, state, transition or trigger.
Relying on UML to model the behavior of an elasticity policy requires to capture correctly the state space of the architecture and the changes in each state. 
There is active work to refine modeling means for application with increased complexity. 
For example, Górski~\cite{10093859} proposes an enhancement of activity diagrams from UML to model the communication between services at the level of sending messages. It is worth investigating how the proposed view type interacts with other view types and viewpoints in other modeling contexts. 

\section{Conclusion and Future Work}

This paper contributes the Slingshot approach that extends Palladio approach with a new view type for modeling the elasticity of the software architecture. We elaborate the new view through the abstract syntax of the SPD modeling language for creating scaling policy definitions, its notation, and its dynamic semantics. 
The collected evidence shows that the proposed view increases Palladio's predictive power by allowing the modeling and analysis of the following styles: centralized host-based policies, decentralized host-based policies, decentralized service-based policies based on application-level metrics, and decentralized service-based policies based on resource-level metrics. 
In two representative cases, it allowed the modeling of 16 different elasticity policies of different styles and configurations that differ in the level of elasticity, control, application awareness, triggering conditions, constraints, type, and magnitude of adjustments.  
The collected empirical evidence through a user study with 12 data points suggests a positive impact on productivity, however, due to the low sample size it is inconclusive. 
When comparing the modeling throughput with the existing empirical evidence, we observe that participants spend more time effectively on an SPD modeling concept than on modeling constructs for control flow, resource environment, or usage profile in Palladio.


For future work, we envision several directions. SPD currently does not support modeling the full spectrum of paradigms for automated resource management that exists in the literature.
For example, we capture the ability to specify proactive policies through anticipatory triggering; that is, we allow the modeler to specify triggers that compute predictive models (e.g., linear regression) on monitored values. 
However, SPD does not allow the modeling of policies that learn. 
A future work direction is to integrate policies with such an attribute into SPD and the simulator for the analysis.
This would allow software architects to model and analyze the full spectrum of approaches to resource management on an architectural level and in composition i.e., employing reactive policies for one portion of the application and policies with learning capabilities for a different portion.   
In addition, the spectrum of approaches also includes adaptation approaches based on other domains, such as control theory or queueing theory. 
Having them all integrated into an architectural-based analysis pipeline would be beneficial for researchers and practitioners. 

When it comes to the analysis itself,  a possible future work item is to improve the tool-support for the analysis of SPD models. 
The evaluation of elasticity policies in terms of quality is semi-automated and involves manual work.
More tool support is needed to map and compare different SPD models with respect to a utility metric. 
Enhancing the analysis in the Palladio ecosystem would lower further the effort of researching adequate policies for various software architectures and architectural patterns. 

\section*{Declaration of generative AI and AI-assisted technologies in the writing process}

During the preparation of this work the author(s) used ChatGPT in order to get proposals on refinements for certain paragraphs. After using this tool/service, the author(s) reviewed and edited the content as needed and take(s) full responsibility for the content of the publication.

\appendix


\printcredits
\bibliographystyle{cas-model2-names}

\bibliography{cas-refs}


\end{document}